\documentclass[aps,prb,reprint,showpacs,amsmath,citeautoscript,
flushbottom,floatfix]{revtex4-1}
\usepackage{graphicx}
\usepackage{dcolumn}
\usepackage{bm}
\usepackage{amsmath}
\usepackage{amssymb}
\setcitestyle{super}

\newcommand{\vc}[1]{\boldsymbol{#1}}

\setcitestyle{numbers,square}

\begin{document}
\title{
Pseudospin exchange interactions in $d^7$ cobalt compounds: Possible
realization of the Kitaev model
}
\author{Huimei Liu and Giniyat Khaliullin}
\affiliation{Max Planck Institute for Solid State Research,
Heisenbergstrasse 1, D-70569 Stuttgart, Germany}

\begin{abstract}
The current efforts to find the materials hosting Kitaev model physics have
been focused on Mott insulators of $d^5$ pseudospin-1/2 ions Ir$^{4+}$ and
Ru$^{3+}$ with $t_{2g}^5$($S=1/2, L=1$) electronic configuration. Here we propose
that the Kitaev model can be realized in materials based on $d^7$ ions
with $t_{2g}^5e_g^2$($S=3/2, L=1$) configuration such as Co$^{2+}$,
which also host the pseudospin-1/2 magnetism. Considering possible
exchange processes, we have derived the $d^7$ pseudospin-1/2 interactions in
90$^{\circ}$ bonding geometry. The obtained Hamiltonian comprises
the bond-directional Kitaev $K$ and isotropic Heisenberg $J$ interactions
as in the case of $d^5$ ions. However, we find that the presence of
additional, spin-active $e_g$ electrons radically changes the balance
between Kitaev and Heisenberg couplings. Most remarkably, we show
that the exchange processes involving $e_g$ spins are highly sensitive
to whether the system is in Mott ($U<\Delta$) or charge-transfer ($U>\Delta$)
insulating regime. In the latter case, to which many cobalt compounds do
actually belong, the antiferromagnetic Heisenberg coupling $J$ is strongly
suppressed and spin-liquid phase can be stabilized. The results suggest
cobalt-based materials as promising candidates for the realization of
the Kitaev model.
\end{abstract}

\date{\today}


\maketitle


\section{Introduction}

In recent years, the Kitaev honeycomb model~\cite{Kit06} and its various
extensions have attracted much attention (see the recent
reviews~\cite{Her17,Kno16,Tre17,Win17,Rau16} and references therein). In this
model, the spins-1/2 interact via a strongly anisotropic, bond-dependent Ising
couplings $S^x_iS^x_{i+\gamma_1}$, $S^y_iS^y_{i+\gamma_2}$, and
$S^z_iS^z_{i+\gamma_3}$, acting on three nearest-neighbor $\gamma_1$,
$\gamma_2$, and $\gamma_3$ bonds of a tri-coordinated honeycomb lattice.
A mutual orthogonality of the Ising-axis directions on different
$\gamma$ bonds results in strong frustration, driving the spins
into a quantum disordered state.

Physically, the bond-dependent exchange couplings as in the Kitaev model may
arise from an unquenched orbital contribution ${\vc L}$ to the magnetic
moments. Due to the non-spherical shape of the electron orbitals, the orbital
moment interactions in transition-metal compounds are strongly anisotropic,
both in real and angular momentum spaces~\cite{Kug82,Kha03,Kha05}.
By virtue of spin-orbit coupling (SOC), this property of orbital interactions
is inherited by total angular momentum ${\vc J}={\vc L}+{\vc S}$ of magnetic
ions, as demonstrated by explicit calculations of magnetic Hamiltonians in the
limit of strong SOC~\cite{Kha05,Che08,Jac09,Cha10}. Apart from magnetism,
SOC driven Kitaev-type interactions may lead also to exotic superconducting
states in doped Mott insulators~\cite{Kha04,Kha05,Hya12,You12}.

From the materials perspective, having unquenched orbital moments ${\vc L}$
in solids is not rare but requires special conditions: (i) lattice
distortions caused by the steric effects are small so that there remains
the orbital (quasi)degeneracy, and (ii) Jahn-Teller coupling and
superexchange interactions, which favor ordering of the real
orbitals and hence quench the ${\vc L}$ moments, are weaker
than intraionic spin-orbit coupling (SOC). Under these conditions, typical for
late transition-metal (TM) compounds with $t_{2g}$ orbital degeneracy,
the exchange interactions between magnetic ions can be formulated in terms
of pseudospins $\widetilde{\vc S}$ operating within the ground-state
spin-orbit manifold. Degeneracy of this manifold, and hence the $\widetilde{S}$
value depends on electron configuration of the TM ions and local crystal
field symmetry.

\begin{figure}
\begin{center}
\includegraphics[width=8.5cm]{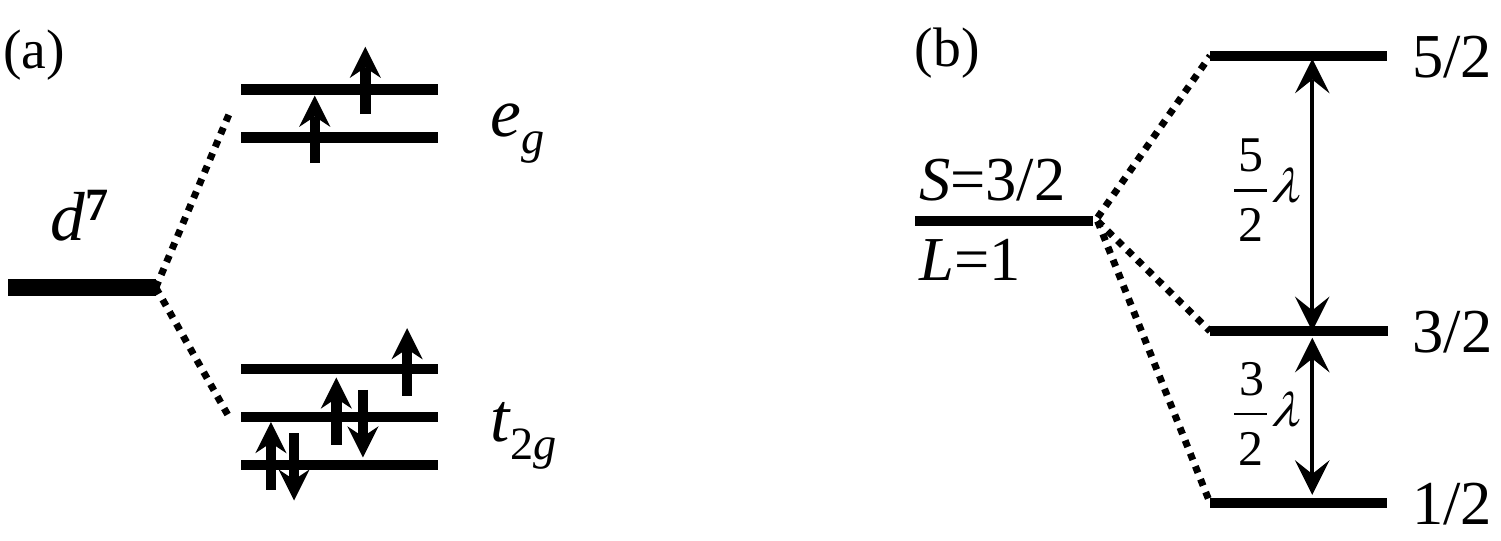}
\caption{
(a) High-spin state of $d^7$($t^5_{2g}e^2_g$) configuration in an octahedral
crystal field.
(b) Splitting of $S=3/2, L=1$ manifold under spin-orbit coupling
$\lambda(\vc L \vc S)$, resulting in pseudospin-1/2 ground state doublet.
}
\label{fig:ene}
\end{center}
\end{figure}

In the context of the Kitaev model, the TM ions that possess pseudospin-1/2
ground state doublet are of particular interest. In the past, $3d$-cobalt
compounds CoO, KCoF$_3$, CoCl$_2$, etc. have been canonical examples of
the pseudospin-1/2 magnetism (see, e.g.,
Refs.~\onlinecite{Hol71,Buy71,Yam74,Gof95}), and, more recently,
$4d$ and $5d$ compounds RuCl$_3$, Sr$_2$IrO$_4$, Na$_2$IrO$_3$, etc. came
into focus (see reviews~\cite{Rau16,Win17}). In cobaltates, the $d^7$ ions
Co$^{2+}$ in an octahedral crystal field have a predominantly $t_{2g}^5e_g^2$
configuration with $S=3/2$ and an effective $L=1$ moments~\cite{Abr70},
forming a pseudospin-1/2 doublet in their ground state
(see Fig.~\ref{fig:ene}). This is similar to the case of $d^5$ ions Ru$^{3+}$
and Ir$^{4+}$ with $t_{2g}^5$($S=1/2$, $L=1$) configuration,
and, on symmetry grounds, the pseudospin-1/2 exchange Hamiltonians in both
$d^7$ and $d^5$ systems must have the identical form. However, the presence
of additional, spin-active $e_g$ electrons in $d^7$ cobaltates is expected
to have a strong impact on the actual values of exchange parameters. In
particular, they should affect the strength of the Kitaev-type couplings
relative to other terms in the Hamiltonian.

In this paper, we derive the $d^7$ pseudospin-1/2 interactions in the
edge-shared, 90$^{\circ}$ bonding geometry, including various nearest-neighbor
hopping processes allowed by symmetry. As expected, the spin-orbital exchange
Hamiltonian projected onto pseudospin $\widetilde{S}=1/2$ subspace comprises
the isotropic Heisenberg $J\widetilde{\bf S}_i\cdot\widetilde{\bf S}_j$ and
bond-dependent Kitaev  $K\widetilde{S}^{\gamma}_i\widetilde{S}^{\gamma}_j$
couplings, as in the case of $d^5$ ions~\cite{Kha05,Che08,Jac09,Cha10}.
(Non-diagonal components of the exchange tensor $\Gamma_{xy}$~\cite{Rau14}
are also present but not dominant.) We find that the presence of $e_g$ spins in
$d^7$ configuration has important consequences on relative values of
the parameters $J$ and $K$. In contrast to $d^5$ case, where the leading
exchange term $\propto 4t^2/U$ does not contribute to the pseudospin
interactions~\cite{Kha05}, we find here that both Heisenberg $J$ and
Kitaev $K$ couplings appear already in the leading order of $4t^2/U$ or
$4t^2/\Delta$ (where $U$ and $\Delta$ stand for intraionic Coulomb and $pd$
charge-transfer energies, correspondingly). Most importantly, the $e_g$ spin
contribution to $J$ is always ferromagnetic and it largely compensates
antiferromagnetic contributions from $t_{2g}$ orbitals. As a result, net value
of isotropic $J$ coupling is strongly reduced, in particular in the
charge-transfer regime of $U>\Delta$ relevant to cobaltates~\cite{Zaa85}.
This mechanism of suppressing the Heisenberg $J$ term is specific to the $d^7$
pseudospin systems, and it suggests an alternative way to access the desired
parameter regime of $K \gg J$ with spin-liquid ground state.

Our study is partially motivated by the recent
experiments~\cite{Vic07,Lef16,Ber17,Won16} on cobalt compounds
Na$_2$Co$_2$TeO$_6$ and Na$_3$Co$_2$SbO$_6$ with a layered hexagonal
structure. In both systems, the $d^7$ ions Co$^{2+}$ form a nearly
perfect honeycomb lattice and develop a zigzag-type antiferromagnetic order
at low temperatures, analogous to that observed in $d^5$ pseudospin-1/2
materials RuCl$_3$ and Na$_2$IrO$_3$. This similarity may not be accidental,
and the results presented in this work suggest that $d^7$ cobalt compounds
may indeed harbor pseudospin-1/2 Kitaev-Heisenberg model and related physics.

This paper is organized as follows. Section~\ref{sec:wavef} introduces the
single ion Hamiltonian and $d^7$ pseudospin-1/2 wave functions.
Section~\ref{sec:hami} discusses various exchange processes between the
$d^7$ TM ions in 90$^{\circ}$ bonding geometry, relevant for a honeycomb
lattice cobaltates, and derives the corresponding exchange parameters as a
function of material parameters. Section~\ref{sec:dis} considers interplay
between different exchange mechanisms and the resulting phase diagrams.
The main results and conclusions are summarized in Sec.~\ref{sec:con}.
Appendix A discusses Hund's coupling corrections to the exchange parameters.

\section{Single ion levels and wave functions}
\label{sec:wavef}

The high-spin $S=3/2$ Co$^{2+}$ ion in the octahedral field has a predominantly
$t_{2g}^5e_g^2$ electronic configuration shown in Fig.~\ref{fig:ene}(a).
(We neglect the admixture of $t_{2g}^4e_g^3$ state, since its spectral weight
is $\sim 0.06$ only~\cite{Pra59,Yam74}.) The threefold orbital
degeneracy of this configuration can conveniently be described in terms of
an effective angular momentum $L=1$~\cite{Abr70}. Further, the spin and
orbital moments are coupled via spin-orbit coupling $\lambda(\vc L\vc S)$,
with $\lambda>0$. This results in a level structure shown
in Fig.~\ref{fig:ene}(b), with the ground-state Kramers doublet hosting
a pseudospin 1/2. The actual value of $\lambda$ is material dependent due to
various factors such as covalency effects, and can experimentally be
quantified from excitation energy $\frac{3}{2}\lambda$ between spin-orbit
levels $1/2 \rightarrow 3/2$ in Fig.~\ref{fig:ene}(b). In Co$^{2+}$ perovskite
KCoF$_3$, this transition (termed as ''spin-orbit exciton'') was observed
at $\sim 40$~meV by the inelastic neutron scattering~\cite{Hol71,Buy71}.

In the cubic crystal field, Co$^{2+}$ pseudospin-1/2 wavefunctions
$|\pm\widetilde{\frac{1}{2}}\rangle$ read, in the basis of $|S_z,L_z\rangle$
states, as follows:
\begin{align}
\Big|\!+\widetilde{\frac{1}{2}}\Big\rangle& =
\frac{1}{\sqrt{2}}\Big|\frac{3}{2},-1\Big\rangle-\frac{1}{\sqrt{3}}\Big|
\frac{1}{2},0\Big\rangle+\frac{1}{\sqrt{6}}\Big|\!-\!\frac{1}{2},1\Big\rangle,
\notag\\
\Big|\!-\widetilde{\frac{1}{2}}\Big\rangle& =
\frac{1}{\sqrt{2}}\Big|\!-\!\frac{3}{2},1\Big\rangle-\frac{1}{\sqrt{3}}\Big|
\!-\!\frac{1}{2},0\Big\rangle+\frac{1}{\sqrt{6}}\Big|\frac{1}{2},-1\Big\rangle.
\label{eq:wf}
\end{align}

As in case of $d^5$ systems~\cite{Kha05}, we derive the $d^7$ pseudospin-1/2
Hamiltonian in the following way: (i) calculate first the exchange
interactions that operate in full spin-orbital Hilbert space, and
(ii) project them onto low-energy pseudospin $\tilde S=1/2$ sector
defined now by the wave functions~\eqref{eq:wf}. The resulting Hamiltonian
does not include the transitions to high-energy states with total angular
momentum 3/2 and 5/2 [Fig.~1(b)]; the corresponding spin-orbit excitations
are assumed to have only perturbative effects on magnetic order and
fluctuations. This approximation, and hence a notion of ''pseudospin'' itself,
is physically justified if the pseudospin interactions are weaker than
spin-orbit coupling. In practice, this criteria implies that the spin-orbit
exciton modes are separated from low-energy pseudospin-1/2 magnons, as
indeed observed in cobaltates KCoF$_3$~\cite{Hol71, Buy71},
GeCo$_2$O$_4$~\cite{Tom11}, NaCaCo$_2$F$_7$~\cite{Ros17}, or in iridate
Sr$_2$IrO$_4$~\cite{Kim12,Kim14}.

The calculations below are based on perturbation theory assuming that
the hopping parameters are smaller than excitation energies of the
intermediate states created during the hopping processes. Specifically,
this implies a hopping amplitude $t_{pd}$ between oxygen $p$ and TM-ion $d$
states is smaller than charge-transfer gap $\Delta$, and hopping $t$ between
TM $d$ orbitals is much smaller than Coulomb $U$ as well as charge-transfer
gap $\Delta$, such that exchange parameters $\propto 4t^2/U$ or
$\propto 4t^2/\Delta$ resulting from calculations are much less than $U$
and $\Delta$ values. Put differently, it is assumed that system is a good
insulator where the magnetic and charge energy scales are well separated,
which is indeed the case in many cobalt compounds with magnon energies
well below 100~meV (see, e.g., Refs.~\cite{Hol71, Buy71,Tom11,Ros17}).

\begin{figure}
\begin{center}
\includegraphics[width=8.5cm]{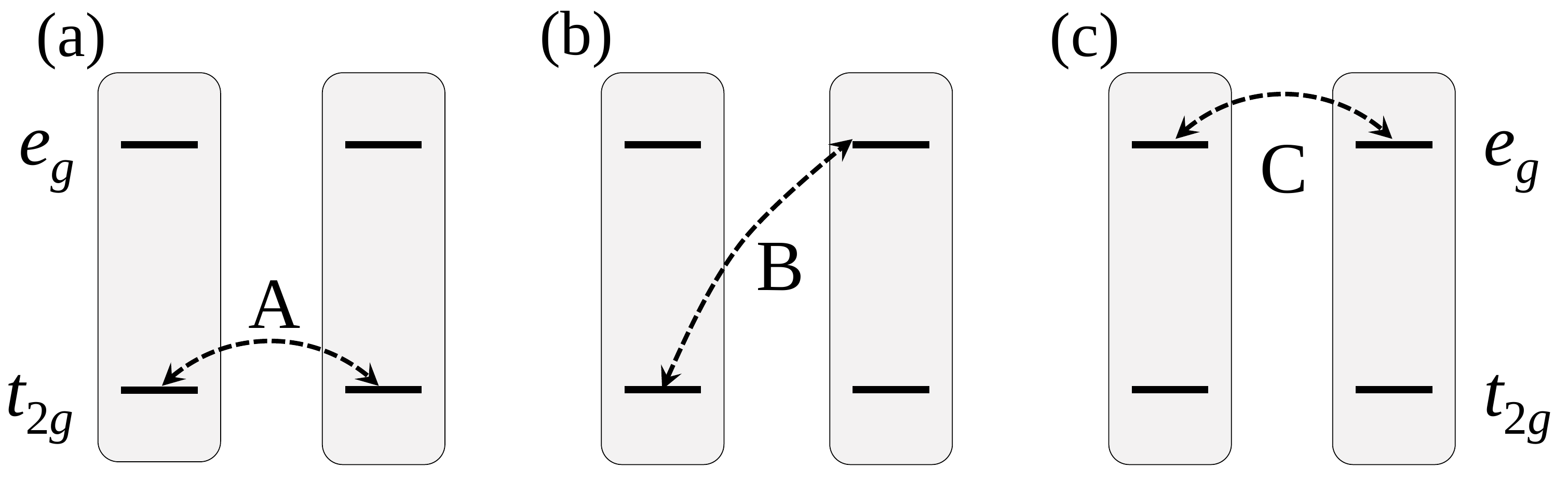}
\caption{Three different classes of the exchange processes, A, B, and C,
derived from interactions between (a) $t_{2g}$ and $t_{2g}$, (b) $t_{2g}$ and
$e_{g}$, and (c) $e_{g}$ and $e_{g}$ spin-orbital levels. In the text, the
corresponding contributions to the exchange Hamiltonian are denoted as
$\mathcal{H}_{\mathrm{A}}$, $\mathcal{H}_{\mathrm{B}}$, and $\mathcal{H}_{\mathrm{C}}$.
}
\label{fig:proc}
\end{center}
\end{figure}

\section{Exchange Processes and Interactions}
\label{sec:hami}

Since the spins residing on $e_{g}$ orbitals also play an active role in
the exchange processes, the spin-orbital model for $d^7$ ions is far more
complex than in $d^5$ systems with $t_{2g}$ only orbitals. To make the
structure of the paper more transparent, we divide the exchange processes
into three classes (see Fig.~\ref{fig:proc}): exchange between
(A) $t_{2g}$ and $t_{2g}$ orbitals, (B) $t_{2g}$ and $e_{g}$ orbitals, and
(C) $e_{g}$ and $e_{g}$ orbitals.

Accordingly, this section is divided into three parts, subsections A, B,
and C, where the above exchange contributions A($t_{2g}$-$t_{2g}$),
B($t_{2g}$-$e_g$), and C($e_g$-$e_g$) are considered. The subsections are
further structured according to three physically distinct exchange
mechanisms: (i) the $U$ processes, (ii) the charge-transfer processes,
and (iii) the cyclic-exchange processes. The $U$ processes involve virtual
excitations with energies of the order of Hubbard $U$ and they are dominant in
Mott-Hubbard-type insulators with $U<\Delta$, while the latter two processes
become important in charge-transfer-type insulators with $\Delta <U$,
in terms of Zaanen-Sawatsky-Allen classification~\cite{Zaa85}. Although
cobalt compounds typically belong to the latter category, we will show the
results for arbitrary values of ratio $U/\Delta$ for the sake of generality.

\subsection{$t_{2g}$-$t_{2g}$ exchange}
\begin{figure}
\begin{center}
\includegraphics[width=8.5cm]{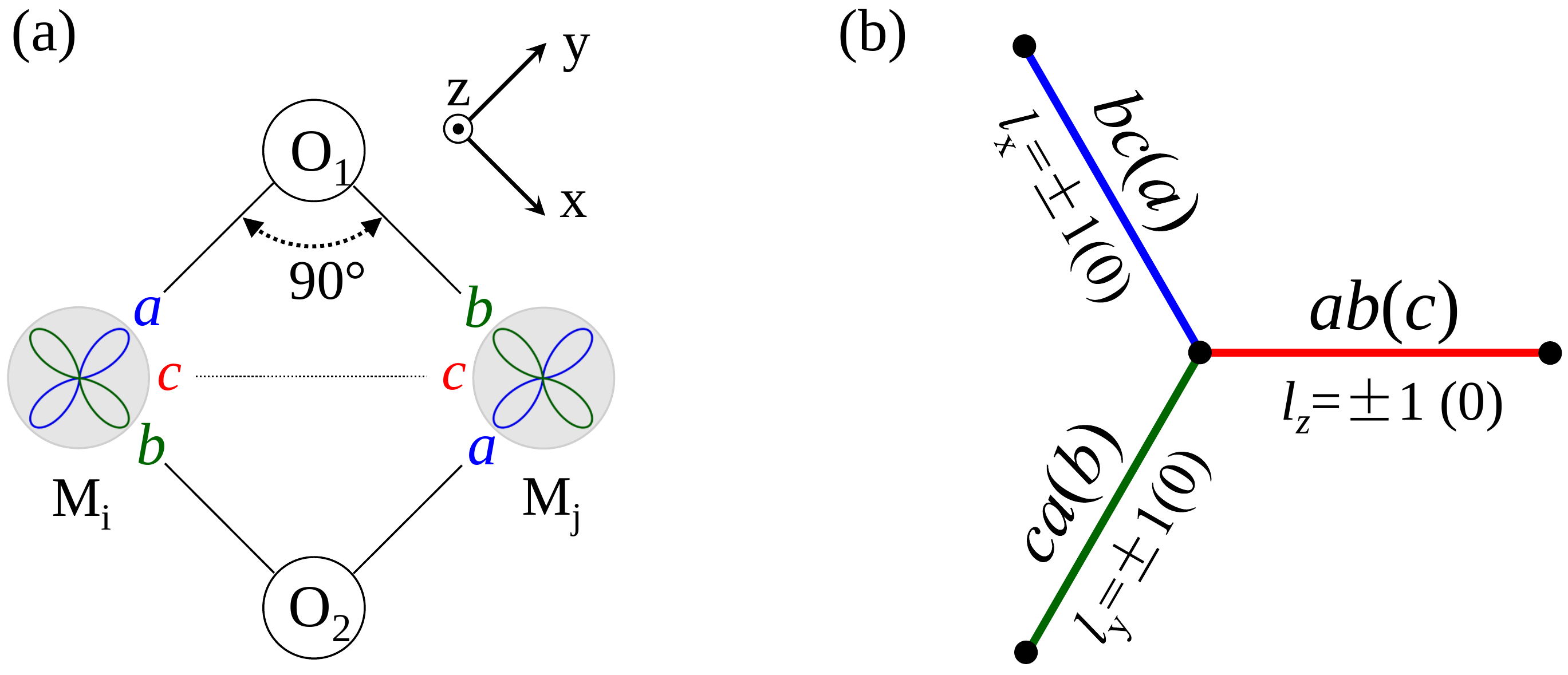}
\caption{(a) A 90$^{\circ}$ M-O-M bonding geometry, where magnetic ions
M$_\mathrm{i}$ and M$_\mathrm{j}$ interact via two oxygen ions O$_1$ and
O$_2$. Two out of three $t_{2g}$ orbitals, here $a = d_{yz}$ and
$b = d_{zx}$, participate in the superexchange process by virtue of
an indirect hopping $t_{ab}$ via the oxygen $p_z$ orbitals.
The remaining orbital $c = d_{xy}$ (not shown) contributes to spin
exchange via a direct hopping $t_{cc}$.
(b) Three types of bonds and corresponding orbital selective hopping geometry.
On the horizontal (red) bond, the $ab(c)$ should read as $t_{ab}$($t_{cc}$),
with $t_{ab} = t$ and $t_{cc} = -t'$. This bond will be referred to as
''$c$ bond'' in the text. In terms of the $t_{2g}$ orbital angular momentum,
$ab$ pair represents the $l_z=\pm 1$ doublet, while $c$ orbital corresponds to
the $l_z=0$ state. The hopping geometry on two other (blue and green) bonds
follows from symmetry.
}
\label{fig:schem}
\end{center}
\end{figure}

\begin{figure}
\begin{center}
\includegraphics[width=8.5cm]{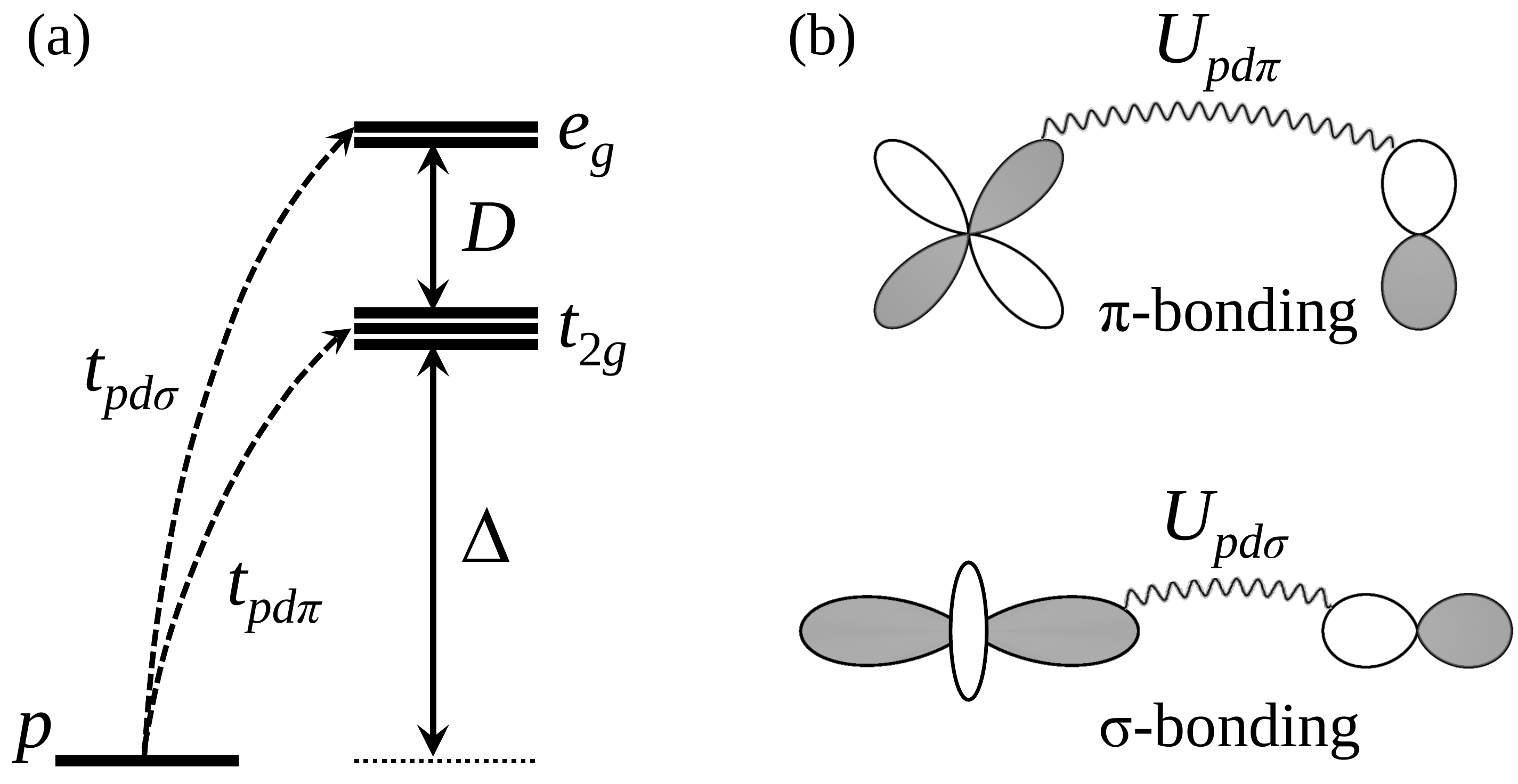}
\caption{(a) Schematic of oxygen $p$ and transition-metal ion
$d(t_{2g},e_g)$ energy levels. $t_{pd\pi}$ ($t_{pd\sigma}$) is the hopping
integral between $p$ and $t_{2g}$ ($e_{g}$) orbitals. The corresponding
$pd$ charge-transfer gaps are $\Delta$ and $\Delta + D$.
(b) Two types of the $pd$-covalent bondings, $\pi$ and $\sigma$,
involving $t_{2g}$ and $e_{g}$ orbitals correspondingly. A $d$ electron and
$p$ hole created by the $pd$ transition experience the Coulomb attraction,
which is stronger in $\sigma$-bonding channel: $U_{pd\sigma}>U_{pd\pi}$. This
reduces the effective value of $D$ from a single-electron cubic splitting
$10Dq$ by $U_{pd\sigma}-U_{pd\pi}>0$ (see text).
}
\label{fig:nota}
\end{center}
\end{figure}

\subsubsection{Intersite $U$ processes}
\label{sec:ttA1}

We consider virtual charge transitions of the type
$d^7_id^7_j \rightarrow d^6_id^8_j$, created by hopping of $t_{2g}$ electrons
between two nearest-neighbor magnetic ions. The excitation energy associated
with this intersite charge fluctuation is Coulomb repulsion $U$, and
the resulting exchange couplings scale as $t^2/U$.

In a 90$^{\circ}$ bonding geometry [see Fig.~\ref{fig:schem}(a)], the
nearest-neighbor $t_{2g}$ orbital hopping along the $c$ bond
can be written as~\cite{Kha04,Kha05,Nor08,Cha11}:
\begin{align}
\mathcal{H}^{(c)}_t = t(a_{i\sigma}^{\dag}b_{j\sigma}+b_{i\sigma}^{\dag}a_{j\sigma})
-t'c_{i\sigma}^{\dag}c_{j\sigma} + \mathrm{H.c}.
\label{eq:tt'}
\end{align}
Here, summation over spin projection $\sigma$ is implied. Parameter
$t= t_{pd\pi}^2/\Delta$ is the hopping amplitude between $a = d_{yz}$ and
$b = d_{zx}$ orbitals, originating from $d$-$p$-$d$ process via the
$p$-$d$ charge-transfer gap $\Delta$ [Fig.~\ref{fig:nota}(a)].
$t'>0$ is given by a direct overlap of $c = d_{xy}$ orbitals.
Note that oxygen-mediated $t$ hopping in Eq.~(\ref{eq:tt'}) changes the orbital
color. In terms of effective angular momentum $l=1$ of $t_{2g}$ electron,
$d_{\pm 1}^{\dag} =\mp (d_{yz}^{\dag} \pm id_{zx}^{\dag})/\sqrt{2}$, this term reads as
$it(d_{1,\sigma}^{\dag}d_{-\!1,\sigma}-d_{-\!1,\sigma}^{\dag}d_{1,\sigma})_{ij}$,
making it clear that hopping does not conserve angular momentum of a pair
and hence may lead to anisotropic exchange interactions.
This is in contrast to 180$^{\circ}$ bonding geometry with orbital-conserving
hopping, $t(a_{i\sigma}^{\dag}a_{j\sigma}+b_{i\sigma}^{\dag}b_{j\sigma})\rightarrow$
$t(d_{1,\sigma}^{\dag}d_{1,\sigma}+d_{-\!1,\sigma}^{\dag}d_{-\!1,\sigma})_{ij}$.

\begin{figure}
\begin{center}
\includegraphics[width=8.5cm]{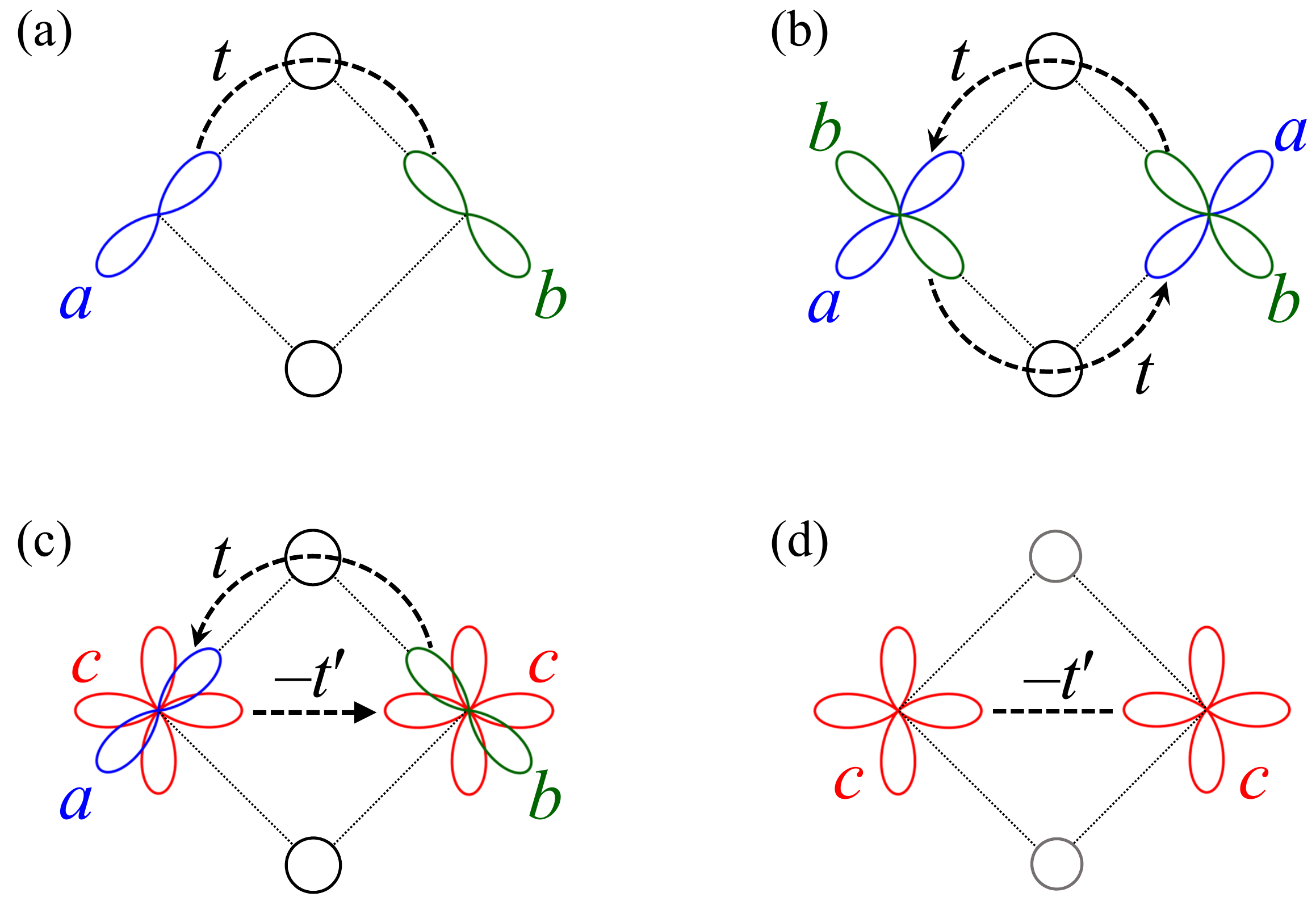}
\caption{
Schematic of the nearest-neighbor hoppings of $t_{2g}$ orbitals $a$ (blue),
$b$ (green), and $c$ (red). Oxygen $p_z$ orbitals are depicted as open circles.
(a) Orbital $a$ to $b$ hopping (and vice versa) through an upper oxygen
$p_z$ orbital.
There is no orbital exchange during this process, and the orbital operator
is $n_{ia}n_{jb}$; similar process through the lower oxygen
gives $n_{ib}n_{ja}$.
(b) The exchange process involving both upper and lower oxygen ions. This
process changes the orbital state $b_ib_j$ into $a_ia_j$. The corresponding
operator is $a_i^{\dag}a_j^{\dag}b_jb_i + \mathrm{H.c.}$
(c) Cross-term involving $t_{ab}$ hopping via oxygen and direct $t'$ hopping
of $c$ orbital.
The orbital exchange operator is $a_i^{\dag}c_j^{\dag}b_jc_i$, and similar
process using the lower oxygen gives $b_i^{\dag}c_j^{\dag}a_jc_i$.
(d) Direct hopping between the $c$ orbitals.
}
\label{fig:t2g1}
\end{center}
\end{figure}

As shown in Fig.~\ref{fig:t2g1} and detailed in its caption, there are several
exchange processes involving different combinations of oxygen-mediated $t$ and
direct $t'$ hoppings. Collecting all these terms, one arrives at the
following spin-orbital Hamiltonian for a pair along $c$ bond:
\begin{align}
\mathcal{H}^{(c)}_{\mathrm{A}1}& =
\frac{4}{9}\frac{t^2}{U}(\vc S_i \cdot \vc S_j+S^2)\left[(n_{ia}n_{jb}+
a_i^{\dag}b_ia_j^{\dag}b_j)+(a\leftrightarrow b)\right]\notag \\
&-\frac{4}{9}\frac{tt'}{U}(\vc S_i \cdot \vc S_j+S^2)
\left[(a_i^{\dag}c_ic_j^{\dag}b_j
+c_i^{\dag}a_ib_j^{\dag}c_j)+(a\leftrightarrow b)\right]\notag \\
&+\frac{4}{9}\frac{t'^2}{U}(\vc S_i \cdot \vc S_j-S^2)\;n_{ic}n_{jc}.
\label{eq:A11}
\end{align}
In this expression, spin $S=3/2$ stands for high-spin configuration of the
$d^7$ ion. We used a relation $\vc s=\frac{1}{2S}\vc S$ between $t_{2g}$-hole
spin $\vc s$ (one-half) and total spin $\vc S$, as dictated by Hund's coupling.
[In principle, the first contribution above contains also a pure density term
$-(t^2/U)(2-n_{ic}-n_{jc})$, which is irrelevant and thus not shown.]

Next, we project the exchange Hamiltonian Eq.~(\ref{eq:A11}) onto
pseudospin $\widetilde{S}=1/2$ subspace defined by wavefunctions
(\ref{eq:wf}). A direct comparison of the matrix elements gives the following
relations:
\begin{align}
S_x &= \tfrac{5}{3}\widetilde{S}_x \;,  \ \ \;
S_x n_a = \tfrac{1}{3}{\widetilde{S}}_x \;,  \ \ \;
S_x n_{b/c} = \tfrac{2}{3}{\widetilde{S}}_x \;, \\
a^{\dag}b &= \tfrac{i}{3}{\widetilde{S}}_z \;,  \ \ \ \ \ \;
b^{\dag}c = \tfrac{i}{3}{\widetilde{S}}_x \;,  \ \ \ \ \ \ \; \;
c^{\dag}a = \tfrac{i}{3}{\widetilde{S}}_y \;,  \\
S_xa^{\dag}b &= \tfrac{1}{6}{\widetilde{S}}_y\;,  \ \ \ \!
S_xb^{\dag}c = \tfrac{5i}{12} \;, \ \ \ \ \ \
S_xc^{\dag}a = \tfrac{1}{6}{\widetilde{S}}_z \;.
\end{align}
The other combinations of spin and orbital operators involving $S_y$ and
$S_z$ can be obtained from the above mappings by symmetry. As a result,
we find the following pseudospin exchange Hamiltonian:
\begin{align}
\mathcal{H}^{(c)}_{\mathrm{A}1}& =
\frac{2}{9}\frac{t^2}{U}\left(\vc{\widetilde{S}}_i \cdot
\vc{\widetilde{S}}_j
-\frac{2}{9}\widetilde{S}^z_i\widetilde{S}^z_j\right)\notag \\
&+\left(\frac{4}{9}\right)^2\frac{tt'}{U}
\left(\widetilde{S}^x_i\widetilde{S}^y_j
+\widetilde{S}^y_i\widetilde{S}^x_j\right)\notag \\
&+\left(\frac{4}{9}\right)^2\frac{t'^2}{U}
\left(\vc{\widetilde{S}}_i \cdot \vc{\widetilde{S}}_j
-\frac{3}{4}\widetilde{S}^z_i\widetilde{S}^z_j\right),
\label{eq:A12}
\end{align}
which comprises antiferromagnetic (AF) Heisenberg
$\vc{\widetilde{S}}_i \cdot \vc{\widetilde{S}}_j$, ferromagnetic (FM)
Kitaev $\widetilde{S}^z_i\widetilde{S}^z_j$, and non-diagonal
$\widetilde{S}^x_i\widetilde{S}^y_j$-type interactions. Interactions
on other ($a$ and $b$) bonds follow from symmetry (cyclic permutations
among $\widetilde{S}^x, \widetilde{S}^y, \widetilde{S}^z$). As already
noticed above, the $U$-process interactions of the order of $t^2/U$ do
not vanish in the present case, in sharp contrast to $d^5$ pseudospin-1/2
systems~\cite{Kha05}. This is due to different internal structure of
pseudospin wave functions in Eq.~(\ref{eq:wf}), as compared to that
of $d^5$ ions with pure $t_{2g}$ orbitals.

\subsubsection{Charge-transfer processes}
\label{sec:ttA2}

\begin{figure}
\begin{center}
\includegraphics[width=8.5cm]{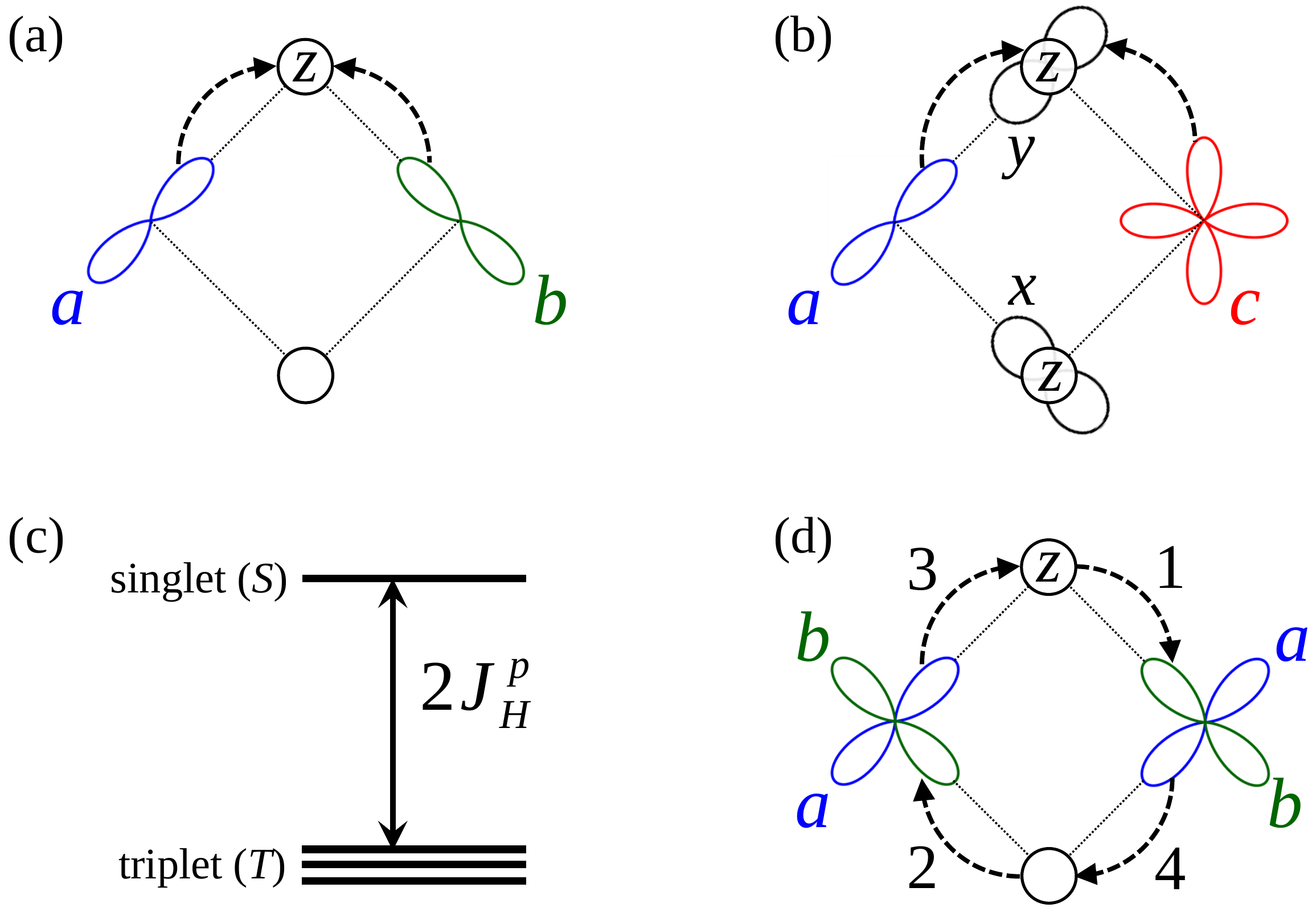}
\caption{(a),(b) Schematic of charge-transfer processes when two holes meet
on an oxygen site. In (a), they arrive at the same $p_z$ orbital, while in (b)
two holes occupy different $p$ orbitals ($z=p_z$ and $y=p_y$) and interact
via Hund's coupling $J^p_H$. The latter results in (c) singlet-triplet
splitting of the intermediate two-hole states of oxygen.
(d) An example of the cyclic exchange process. The numbers 1, 2, 3, 4 represent
the time order of the $pd$ hoppings of electrons. During this process, the
particles cycle within the plaquette avoiding a direct contact with each
other. The ions interchange their spin and orbital quantum numbers, and this
can be expressed as a product of Dirac spin permutation operator and
orbital operator $b_i^{\dag}b_j^{\dag}a_ja_i$.}
\label{fig:t2g2}
\end{center}
\end{figure}

We consider now virtual $pd$ charge-transfer excitations of the type
$d^7_i\!-\!p^6\!-\!d^7_j \rightarrow d^8_i\!-\!p^4\!-\!d^8_j$,
when two holes are created on
an oxygen site bridging nearest-neighbor magnetic ions $i$ and $j$, see
Figs.~\ref{fig:t2g2}(a) and \ref{fig:t2g2}(b). If the holes meet at the same $p$ orbital,
intermediate-state energy is $2\Delta + U_p$, where $U_p$ is Coulomb
repulsion on oxygen site. This contribution is antiferromagnetic, since two
holes have to be in spin-singlet state. If the holes occupy different
orbitals [e.g., $p_z$ and $p_y$ as in Fig.~\ref{fig:t2g2}(b)], the intermediate-state 
energy depends on whether the two holes form triplet ($T$) or singlet
($S$) states, with $E_T = 2\Delta+U'_p-J^p_H$ and $E_S = 2\Delta+U'_p+J^p_H$,
where $U'_p=U_p-2J^p_H$, and $J^p_H$ is Hund's coupling on oxygen that
splits $T$ and $S$ states [see Fig.~\ref{fig:t2g2}(c)]. The exchange energy gain
in this case is
\begin{align}
-4t^2\left(\frac{1}{E_T}P_T+\frac{1}{E_S}P_S\right),
\end{align}
where
$P_T=\tfrac{3}{4}+(\vc s_i \cdot \vc s_j)$ and
$P_S=\tfrac{1}{4}-(\vc s_i \cdot \vc s_j)$
are triplet- and singlet-state projectors. Since $E_T<E_S$, this contribution is
of a ferromagnetic nature.

Collecting the above charge-transfer contributions, we obtain the following
spin-orbital Hamiltonian for a $c$-bond pair:
\begin{align}
\mathcal{H}^{(c)}_{\mathrm{A}2}& = \frac{4}{9}\frac{t^2}{\Delta+\frac{U_p}{2}}
(\vc S_i \cdot \vc S_j-S^2)(n_{ia}n_{jb}+n_{ib}n_{ja})\notag \\
&-\frac{2}{9}\frac{t^2 \; J^p_H}{(\Delta+\frac{U'_p}{2})^2}\;
\vc S_i \cdot \vc S_j(n_{ic}+n_{jc}).
\label{eq:A2}
\end{align}
Projecting this Hamiltonian onto the ground-state doublet (\ref{eq:wf}),
we find:
\begin{align}
\mathcal{H}^{(c)}_{\mathrm{A}2}& = \frac{16}{81}\frac{t^2}{\Delta+\frac{U_p}{2}}
\left(\vc{\widetilde{S}}_i \cdot \vc{\widetilde{S}}_j
+\widetilde{S}^z_i\widetilde{S}^z_j\right)\notag \\
&-\frac{40}{81}\frac{t^2 \; J^p_H}{(\Delta+\frac{U'_p}{2})^2}\;
\left(\vc{\widetilde{S}}_i \cdot \vc{\widetilde{S}}_j
-\frac{1}{2}\widetilde{S}^z_i\widetilde{S}^z_j\right).
\end{align}
It follows that charge-transfer contribution to the Kitaev term is of
a positive sign (i.e., AF), while Heisenberg coupling can be either AF or FM
depending on Hund's coupling strength $J^p_H$.

\subsubsection{Cyclic exchange processes}
\label{sec:ttA3}

The nearest-neighbor cyclic exchange process [see Fig.~\ref{fig:t2g2}(d) and
its caption] is special to 90$^{\circ}$ bonding geometry. This exchange
involves two oxygen sites (O$_1$ and O$_2$) where two holes are generated
in the intermediate state, i.e., $d^7_i\!-\!(p_1^6,p_2^6)\!-\!d^7_j \rightarrow
d^8_i\!-\!(p_1^5,p_2^5)\!-\!d^8_j$. This process is distinct from the above
$U$ and $2\Delta+U_p$ processes in a sense that the excited two particles
do not meet either at TM or oxygen ions, and thus no direct Coulomb repulsion
is encountered. Exchange interaction in this case is of a pure
quantum-mechanical origin: during the cyclic motion of the electrons,
two TM ions interchange their spin and orbital quantum numbers, and the
resulting kinetic energy gain depends on symmetry of a wave function and
hence on total spin of a pair. The cyclic exchange Hamiltonian can be expressed
via Dirac spin permutation operator $(2\vc s_i \cdot \vc s_j+\tfrac{1}{2})
\rightarrow (\vc S_i \cdot \vc S_j+S^2)/2S^2$ and the orbital exchange
operators as follows:
\begin{equation}
\mathcal{H}^{(c)}_{\mathrm{A}3} = \frac{4}{9}\frac{t^2}{\Delta}(\vc S_i \cdot
\vc S_j+S^2)(a_i^{\dag}b_ia_j^{\dag}b_j+b_i^{\dag}a_ib_j^{\dag}a_j).
\label{eq:A3}
\end{equation}
After projecting onto a pseudospin-1/2 doublet (\ref{eq:wf}), this
Hamiltonian reads as follows:
\begin{equation}
\mathcal{H}^{(c)}_{\mathrm{A}3} = \frac{2}{81}\frac{t^2}{\Delta}
(\vc{\widetilde{S}}_i \cdot \vc{\widetilde{S}}_j
-10\widetilde{S}^z_{i}\widetilde{S}^z_j).
\end{equation}
It follows from this equation that the cyclic exchange leads to nearly pure
Kitaev coupling with $K=-10J$. Taken alone, this exchange mechanism would
lead to the spin-liquid ground state (which is stable for
$-K>8J$~\cite{Cha10,Cha13}).

Now, we put together all three $t_{2g}-t_{2g}$ exchange contributions
considered in this subsection, and obtain
\begin{align}
\mathcal{H}^{(c)}_{\mathrm{A}}& = \mathcal{H}^{(c)}_{\mathrm{A}1}
+\mathcal{H}^{(c)}_{\mathrm{A}2}+\mathcal{H}^{(c)}_{\mathrm{A}3} \notag \\
&= J_\mathrm{A}\vc{\widetilde{S}}_i \cdot \vc{\widetilde{S}}_j
+K_\mathrm{A}\widetilde{S}^z_i\widetilde{S}^z_j
+\Gamma_\mathrm{A}(\widetilde{S}^x_i\widetilde{S}^y_j
+\widetilde{S}^y_i\widetilde{S}^x_j),
\end{align}
with the following parameters:
\begin{align}
J_{\mathrm{A}}& \!= \!+\frac{2}{9}t^2\!\left[(1\!+\!\frac{8\kappa^2}{9})
\frac{1}{U}\!+\!\frac{8}{9}\frac{1}{\Delta\!+\!\frac{U_p}{2}}\!-\!
\frac{20}{9}\frac{J^p_H}{(\Delta\!+\!\frac{U'_p}{2})^2}\!+\!
\frac{1}{9\Delta}\right]\!, \notag \\
K_{\mathrm{A}}& \!=
\!-\frac{2}{9}t^2\!\left[(\frac{2}{9}\!+\!\frac{2\kappa^2}{3})
\frac{1}{U}\!-\!\frac{8}{9}\frac{1}{\Delta\!+\!\frac{U_p}{2}}\!-\!
\frac{10}{9}\frac{J^p_H}{(\Delta\!+\!\frac{U'_p}{2})^2}\!+\!
\frac{10}{9\Delta}\right]\!, \notag \\
\Gamma_{\mathrm{A}}& \!= \frac{2}{9}t^2\;\frac{8\kappa}{9U}\;.
\label{eq:JKA}
\end{align}
Here, a ratio $\kappa = t'/t$ between a direct $t'$ and oxygen-mediated $t$
hoppings (depending on material chemistry) is introduced. Typically,
$\kappa<1$ for $3d$-ion wave functions, which implies that non-diagonal
component of the exchange tensor $\Gamma$ is small.

\begin{figure}
\begin{center}
\includegraphics[width=8.5cm]{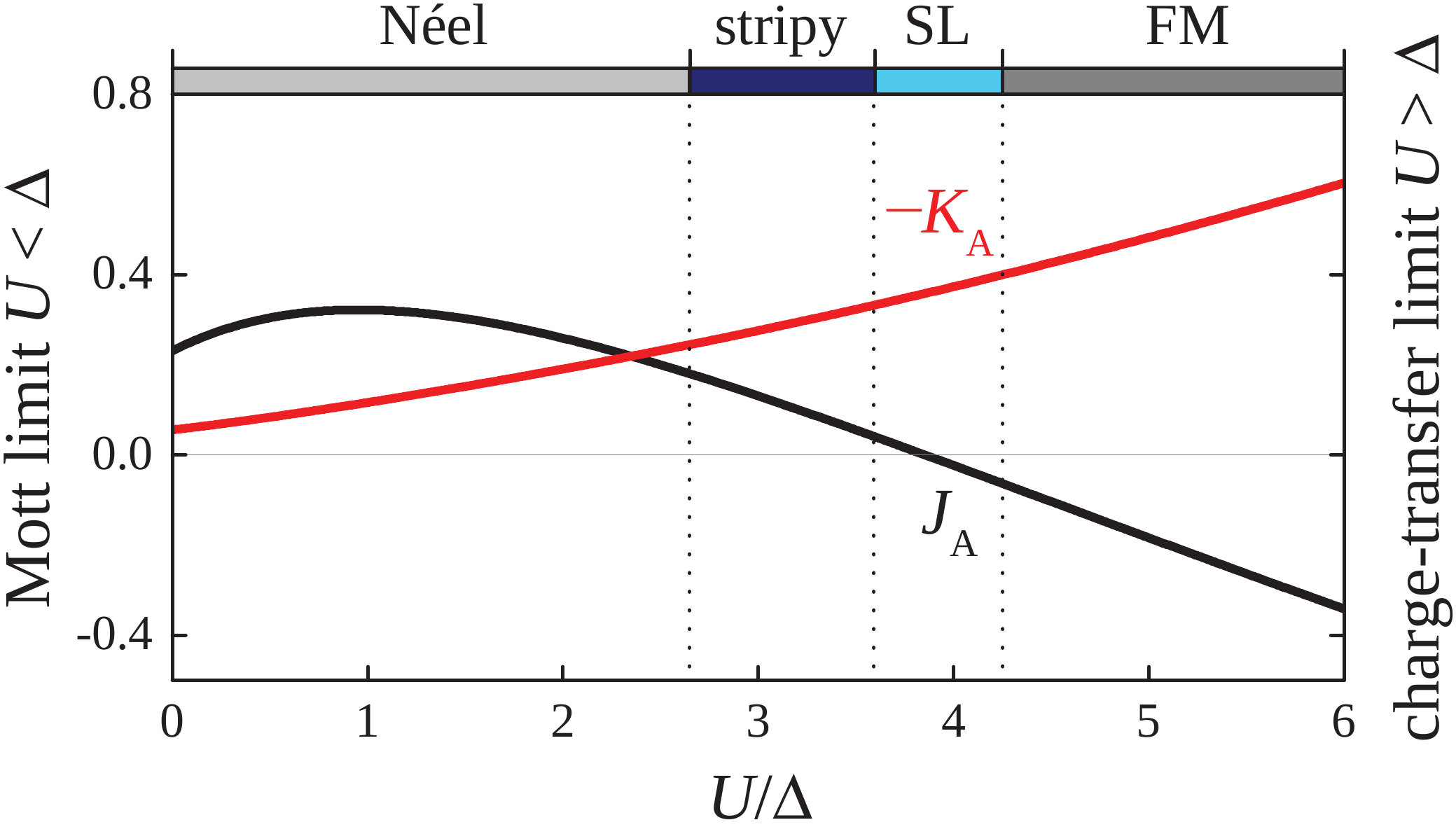}
\caption{$t_{2g}$-$t_{2g}$ contribution to Heisenberg ($J_{\mathrm{A}}$) and
Kitaev ($-K_{\mathrm{A}}$) couplings in units of $t^2/U$ as a function of
$U/\Delta$. The parameters used are: $J^p_H/U_p$=0.3 \cite{Foy13},
$\kappa$=0.2, and $U_p/U$=0.7. The phase boundaries are obtained using the
results of Refs.~\onlinecite{Cha10,Cha13}. SL is the abbreviation for spin
liquid state. In the ''Mott limit'' of $\Delta \gg U$, interactions are
dominated by the $U$ processes $\propto t^2/U$.}
\label{fig:kaja}
\end{center}
\end{figure}

Regarding Heisenberg $J$ and Kitaev $K$ couplings, it follows from
Eq.~(\ref{eq:JKA}) that their ratio strongly depends on whether the system
is in Mott ($U<\Delta$) or charge-transfer ($U>\Delta$) insulating
regime~\cite{Zaa85}. In the first case, the $U$ processes dominate resulting in
$J_A>-K_A$. This ratio is reversed when charge-transfer and cyclic exchange
processes start to dominate as $U/\Delta$ increases. Figure~\ref{fig:kaja},
where we plot $J_A$ and $K_A$ values as a function of $U/\Delta$, clearly
illustrates this trend. For $\kappa=0.2$ used in this figure, parameter
$\Gamma_A$ is small, $\Gamma_A \ll (J_A, K_A)$, and not shown.

The phase diagram of the Kitaev-Heisenberg model as a function of $K/J$ has
been quantified in Refs.~\onlinecite{Cha10,Cha13}. Using the results of these
works, we have indicated in Fig.~\ref{fig:kaja} the phase boundaries.
In the Mott insulating regime of small $U/\Delta$ (i.e., $\Delta \gg U \gg t$),
N\'{e}el and stripy AF states are stable. The spin-liquid state with
$-K_A \gg |J_A|$ is expected around $U/\Delta\sim 4$.
In strong charge-transfer limit ($U \gg \Delta \gg t$), this state
gives way to the FM phase.

\subsection{$t_{2g}$-$e_g$ exchange}

In 90$^{\circ}$ bonding geometry, hopping between $t_{2g}$ and $e_g$ orbitals
is actually the largest one, since it involves $\sigma$-type
$t_{pd\sigma} (\simeq 2t_{pd\pi})$ overlap. Therefore, the $t_{2g}$-$e_g$
exchange contributions to $J$ and $K$ couplings are essential.
In this subsection, we quantify these contributions. Technically, the
calculations closely follow those in previous section, with only
difference arising from different geometry of the orbitals involved.

\subsubsection{Intersite $U$ processes}
\label{sec:teB1}

For a $c$-bond pair, hopping between $c$ and $3z^2-r^2$-type orbitals, shown
in Fig.~\ref{fig:t2geg}(a), is only finite~\cite{Kha05,Cha08}; the other
$t_{2g}$-$e_g$ matrix elements vanish due to quantum interference between
$pd$ virtual hoppings through the upper and lower oxygen ions. Since $e_g$
states are half-filled in $d^7$ configuration, the exchange interaction should
be antiferromagnetic, according to Goodenough-Kanamori rules~\cite{Goo63}.
Explicit calculations of the energy gain from hopping processes in
Fig.~\ref{fig:t2geg}(a) indeed result in AF spin coupling
\begin{equation}
\mathcal{H}^{(c)}_{\mathrm{B}1} = \frac{4\alpha}{9}\frac{tt_e}{\widetilde{U}}
(\vc S_i \cdot \vc S_j-S^2)(n_{ic}+n_{jc}).
\label{eq:B1}
\end{equation}
Here, we introduced $t_e= t_{pd\sigma}^2/\Delta_e$, and
\begin{align}
&\alpha = 1-\frac{D^2}{2\Delta\Delta_e}
\left(\frac{\Delta+\Delta_e}{U}-1\right),\notag \\
&\frac{1}{\widetilde{U}} = \frac{1}{2}\left(\frac{1}{U+D}+\frac{1}{U-D}\right).
\end{align}
In these equations, $\Delta_e=\Delta+D$ stands for $p \rightarrow e_g$
charge-transfer energy, see Fig.~\ref{fig:nota}(a). Parameter $D$ represents
a difference between the $pd$ charge-transfer gaps for $t_{2g}$ and
$e_{g}$ states; we note that it is actually smaller than a single-electron
cubic splitting $10Dq$ due to excitonic effects. Namely, $pd$ excitation
energy $\Delta_{pd}$ is renormalized by electron-hole attraction:
$\Delta_{pd} \rightarrow (E_d-E_p)-U_{pd}$. Due to different spatial shapes of
the orbitals, involved in $\pi$- and $\sigma$-type bondings
[see Fig.~\ref{fig:nota}(b)], one has $U_{pd\sigma}>U_{pd\pi}$. As a result,
\begin{align}
&D= 10Dq-\delta U_{pd}, \ \ \text{with}\ \ \delta U_{pd}=U_{pd\sigma}-U_{pd\pi}.
\label{eq:D}
\end{align}

The exchange Hamiltonian (\ref{eq:B1}), projected onto pseudospin-1/2 sector
reads as:
\begin{equation}
\mathcal{H}^{(c)}_{\mathrm{B}1} = \frac{80\alpha}{81}\frac{tt_e}{\widetilde{U}}
\left(\vc{\widetilde{S}}_i \cdot \vc{\widetilde{S}}_j-
\frac{1}{2}\widetilde{S}^z_i\widetilde{S}^z_j\right),
\label{eq:b1}
\end{equation}
comprising AF Heisenberg and FM Kitaev terms.

\subsubsection{Charge-transfer processes}
\label{sec:teB2}

\begin{figure}
\begin{center}
\includegraphics[width=8.5cm]{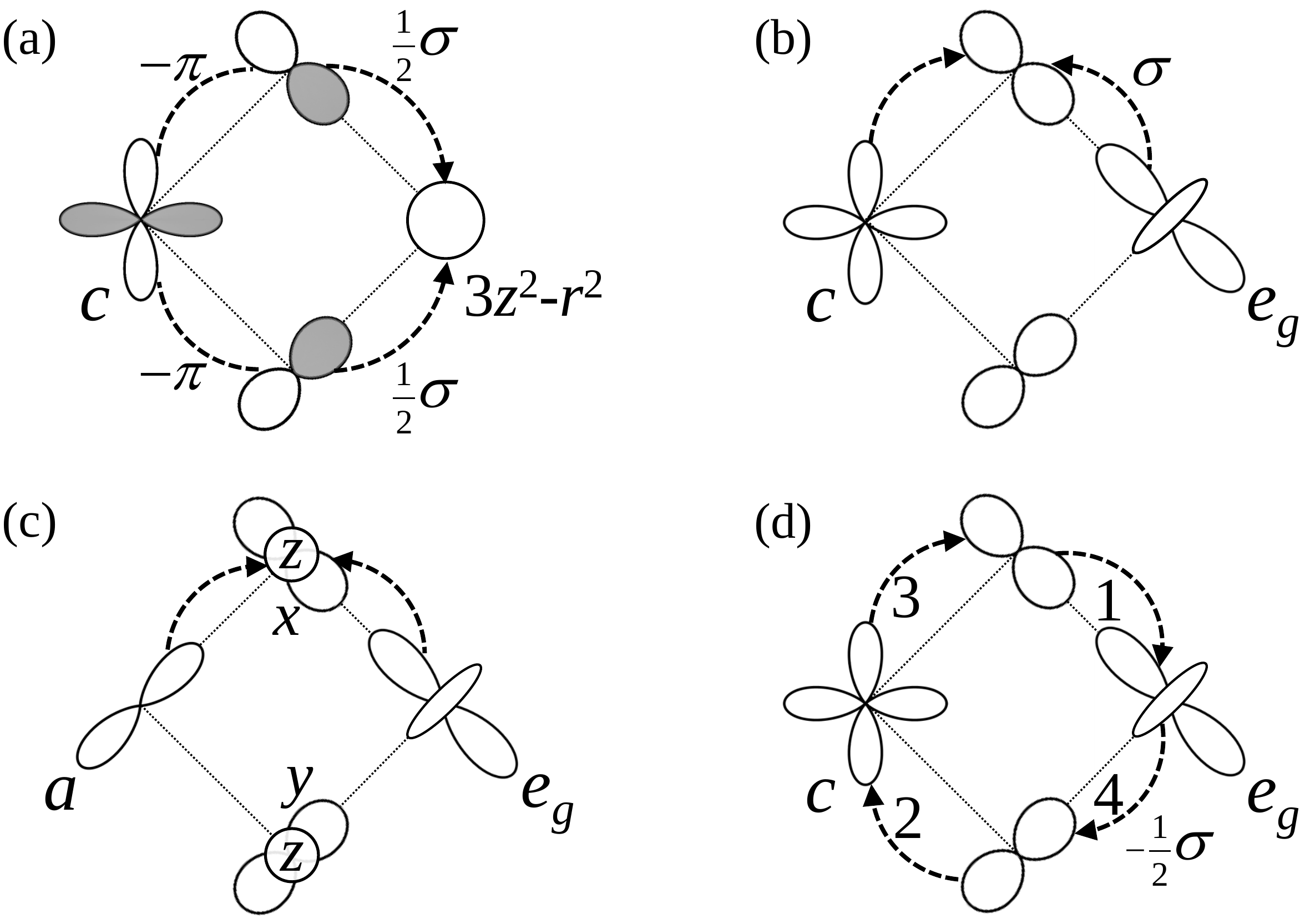}
\caption{Schematic of $t_{2g}-e_g$ hopping processes. $\pi$ ($\sigma$)
represents the $t_{pd\pi}$ ($t_{pd\sigma}$) overlap between $t_{2g}$ ($e_g$)
and $p$ orbitals.
(a) Intersite $U$ process resulting from hopping of $c=d_{xy}$ orbital to
$e_g(3z^2-r^2)$ orbital. In this panel, the shaded (blank) lobes of
wavefunctions imply a positive (negative) sign.
(b), (c) Charge-transfer processes when two holes meet each other at an oxygen
ion, and occupy either (b) the same $p$ orbital, or (c) different $p$
orbitals and interact via Hund's coupling $J^p_H$.
(d) An example of the cyclic exchange process. The numbers 1,2,3,4 indicate
the time order of the $pd$-hoppings of electrons. The overlap between $e_g$
orbital and lower oxygen $p_y$ orbital obtains a prefactor (-1/2).
}
\label{fig:t2geg}
\end{center}
\end{figure}

There are two distinct charge-transfer contributions involving $t_{2g}$ and
$e_g$ orbitals, see Figs.~\ref{fig:t2geg}(b) and (c).
Calculations similar to those in subsection-A2 above
result in the spin-orbital Hamiltonian:
\begin{align}
\mathcal{H}^{(c)}_{\mathrm{B}2}& =
\frac{8\beta}{9}\frac{tt_e}{\Delta+\frac{U_p}{2}}(\vc S_i \cdot \vc S_j-S^2)
(n_{ic}+n_{jc})\notag \\
&-\frac{2\gamma}{9}\frac{tt_e \; J^p_H}{(\Delta+\frac{D+U'_p}{2})^2}\;
\vc S_i \cdot \vc S_j(n^i_{ab}+n^j_{ab}).
\label{eq:b2}
\end{align}
Here, $n_{ab}=n_a+n_b$, and parameters
\begin{align}
\beta& = 1-\frac{D}{4(\Delta_e+\frac{U_p}{2})}
+\frac{D\;U_p}{8\Delta(\Delta_e+\frac{U_p}{2})}-\frac{D}{4\Delta_e},\notag
\\
\gamma& =\frac{(\Delta+\Delta_e)^2}{4\Delta\Delta_e}\;.
\end{align}

After projecting Eq.~(\ref{eq:b2}) onto the Kramers doublet, one obtains the
pseudospin-1/2 exchange Hamiltonian:
\begin{align}
\mathcal{H}^{(c)}_{\mathrm{B}2}& =
\frac{20\beta}{27}\;\frac{8}{3}\frac{tt_e}{\Delta
+\frac{U_p}{2}}\left(\vc{\widetilde{S}}_i \cdot \vc{\widetilde{S}}_j
-\frac{1}{2}\widetilde{S}^z_i\widetilde{S}^z_j\right)\notag \\
&-\frac{20\gamma}{27}\frac{tt_e \; J^p_H}{(\Delta+\frac{D+U'_p}{2})^2}
\left(\vc{\widetilde{S}}_i \cdot \vc{\widetilde{S}}_j
+\frac{1}{3}\widetilde{S}^z_i\widetilde{S}^z_j\right).
\end{align}

\subsubsection{Cyclic exchange processes}
\label{sec:teB3}

This process involves $c$ and $e_g$ orbitals as depicted in
Fig.~\ref{fig:t2geg}(d). As discussed in subsection-A3, the cyclic
exchange Hamiltonian can be expressed via Dirac spin permutation operator,
and orbital projector $n_c$ (for a $c$ bond):
\begin{equation}
\mathcal{H}^{(c)}_{\mathrm{B}3}=-\frac{2\zeta}{9}\frac{tt_e}{\Delta}
(\vc S_i \cdot \vc S_j+S^2)(n_{ic}+n_{ic}),
\label{eq:B3}
\end{equation}
with
\begin{equation}
\zeta=1-\frac{1}{2}\frac{D}{\Delta+D}.
\end{equation}
As compared to $t_{2g}-t_{2g}$ cyclic exchange (\ref{eq:A3}), an overall
negative sign appears in Eq.~(\ref{eq:B3}); it originates from $p_y$-$e_g$
overlap phase factor ($-1/2$) indicated in Fig.~\ref{fig:t2geg}(d).

The corresponding pseudospin-1/2 exchange Hamiltonian is:
\begin{equation}
\mathcal{H}^{(c)}_{\mathrm{B}3}=-\frac{40\zeta}{81}\frac{tt_e}{\Delta}
\left(\vc{\widetilde{S}}_i \cdot \vc{\widetilde{S}}_j
-\frac{1}{2}\widetilde{S}^z_i\widetilde{S}^z_j\right).
\end{equation}

Putting now together all the three $t_{2g}-e_g$ contributions above,
$\mathcal{H}^{(c)}_{\mathrm{B}1} + \mathcal{H}^{(c)}_{\mathrm{B}2}
+ \mathcal{H}^{(c)}_{\mathrm{B}3}$, we get:
\begin{align}
\mathcal{H}^{(c)}_{\mathrm{B}} =
J_{\mathrm{B}}\vc{\widetilde{S}}_i \cdot \vc{\widetilde{S}}_j
+K_{\mathrm{B}}\widetilde{S}^z_i\widetilde{S}^z_j\;,
\end{align}
with
\begin{align}
J_{\mathrm{B}}& = +\frac{80}{81}tt_e\!\left[\frac{\alpha}{\widetilde{U}}
+\frac{2\beta}{\Delta\!+\!\frac{U_p}{2}}-\frac{3\gamma}{4}
\frac{J^p_H}{(\Delta\!+\!\frac{D+U'_p}{2})^2}-\frac{\zeta}{2\Delta}\right]\!,
\notag \\
K_{\mathrm{B}}& = -\frac{40}{81}tt_e\!\left[\frac{\alpha}{\widetilde{U}}
+\frac{2\beta}{\Delta\!+\!\frac{U_p}{2}}+\;\frac{\gamma}{2}
\frac{J^p_H}{(\Delta\!+\!\frac{D+U'_p}{2})^2}-\frac{\zeta}{2\Delta}\right]\!.
\label{eq:JKB}
\notag \\
\end{align}

\begin{figure}
\begin{center}
\includegraphics[width=8.5cm]{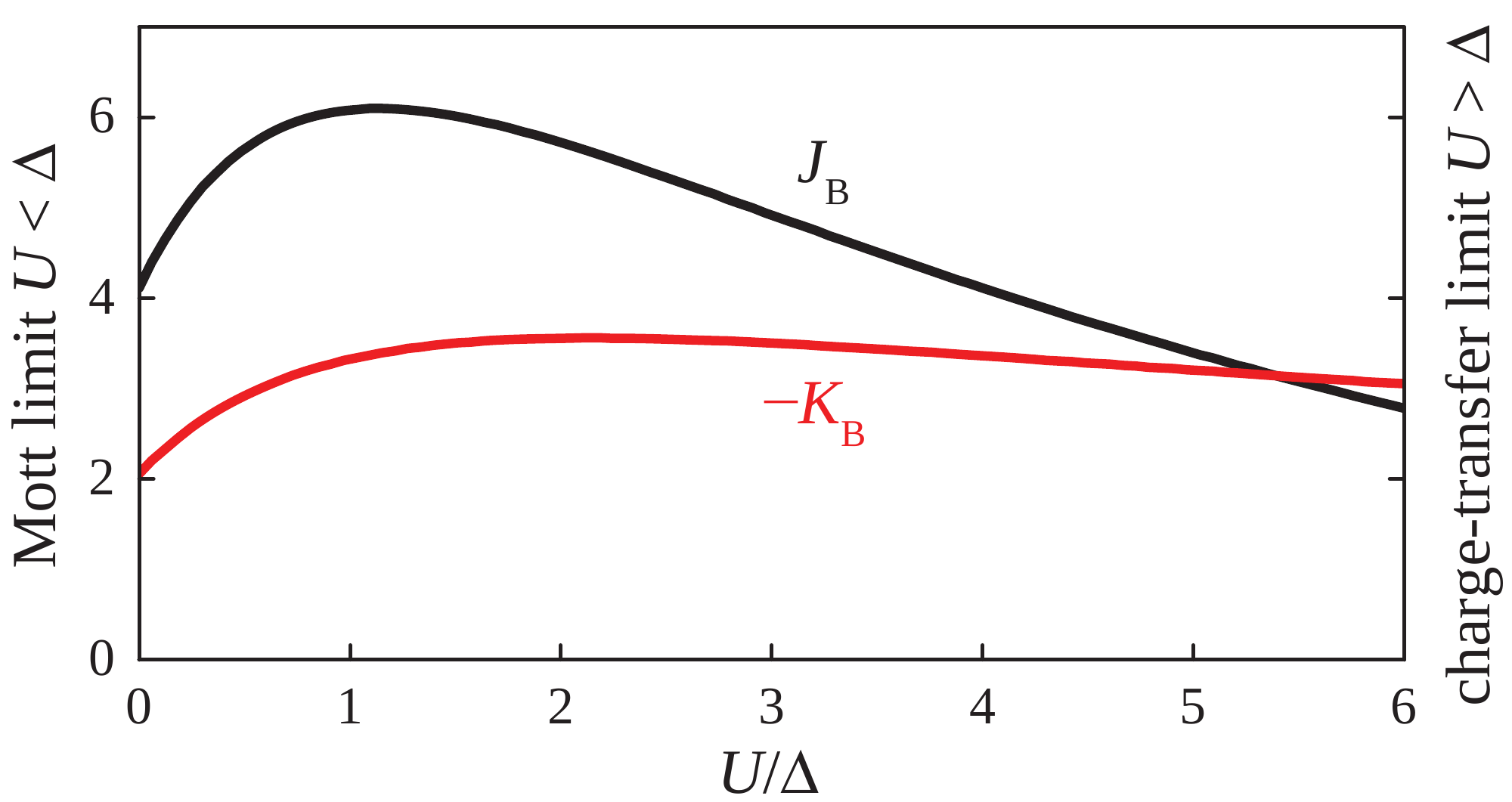}
\caption{$t_{2g}$-$e_g$ contribution to Heisenberg ($J_{\mathrm{B}}$) and
Kitaev ($-K_{\mathrm{B}}$) couplings in units of $t^2/U$ as a function
of $U/\Delta$. The parameters used are: $J^p_H/U_p$=0.3~\cite{Foy13},
$t_{pd\sigma}/t_{pd\pi}=2$, $D/U=0.2$, and $U_p/U=0.7$.}
\label{fig:kbjb}
\end{center}
\end{figure}

In Fig.~\ref{fig:kbjb}, we show how Heisenberg ($J_{\mathrm{B}}$) and Kitaev
($-K_{\mathrm{B}}$) couplings vary as a function of $U/\Delta$.
As in the case of $t_{2g}-t_{2g}$ exchange, $J_{\mathrm{B}}$ dominates in Mott
limit. When $U/\Delta$ increases, the charge-transfer and cyclic exchange
contributions gradually increase, resulting in comparable values of FM Kitaev
and AF Heisenberg couplings.

Comparing the overall values of $t_{2g}-t_{2g}$ and $t_{2g}-e_g$ exchange
contributions, represented in Figs.~\ref{fig:kaja} and \ref{fig:kbjb}
correspondingly, one immediately notices the dominance of the $t_{2g}-e_g$
exchange channel, as expected on general grounds: as noticed above,
$t_{2g}-e_g$ hopping is the largest one in 90$^{\circ}$ bonding
geometry~\cite{Kha05,Cha08}, and the $d^7$ pseudospin-1/2 wavefunction
contains a large weight of the $e_g$-level spin density.

\subsection{$e_g$-$e_g$ exchange processes}
\label{sec:ee}

\begin{figure}
\begin{center}
\includegraphics[width=8.5cm]{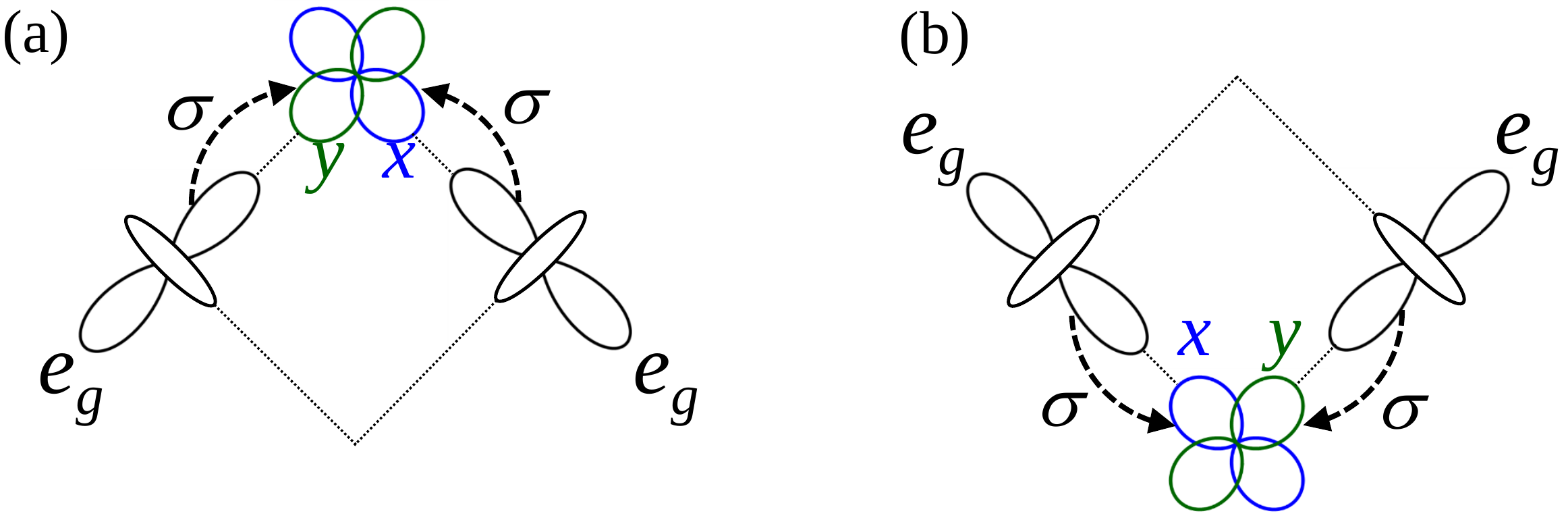}
\caption{Schematic of the $e_g-e_g$ charge-transfer exchange process leading
  to ferromagnetic Heisenberg coupling. Two holes meet on (a) upper
  oxygen or (b) lower oxygen ions, and interact via Hund's coupling $J^p_H$.
  $\sigma$ stands for $t_{pd\sigma}$ hopping between $e_g$ orbitals and
  $p_x$ (blue) and $p_y$ (green) orbitals.
}
\label{fig:eg}
\end{center}
\end{figure}

Finally, we consider the pseudospin interactions originating from
nearest-neighbor coupling of the spins residing on $e_g$ orbitals, see
Fig.~\ref{fig:proc}(c). In 90$^{\circ}$ bonding geometry, both $U$ and
cyclic-exchange processes of $e_g$ spins vanish by symmetry, and we are left
with the charge-transfer process alone, where two holes are transferred to an
oxygen ion and interact via Hund's coupling $J^p_H$, see Fig.~\ref{fig:eg}.
Similar to the $t_{2g}-t_{2g}$ charge-transfer process in
Fig.~\ref{fig:t2g2}(b), this contribution gives FM coupling, as expected from
Goodenough-Kanamori rules~\cite{Goo63} for orbitals that do not directly
overlap and interact via Hund's coupling.

\begin{figure}
\begin{center}
\includegraphics[width=8.5cm]{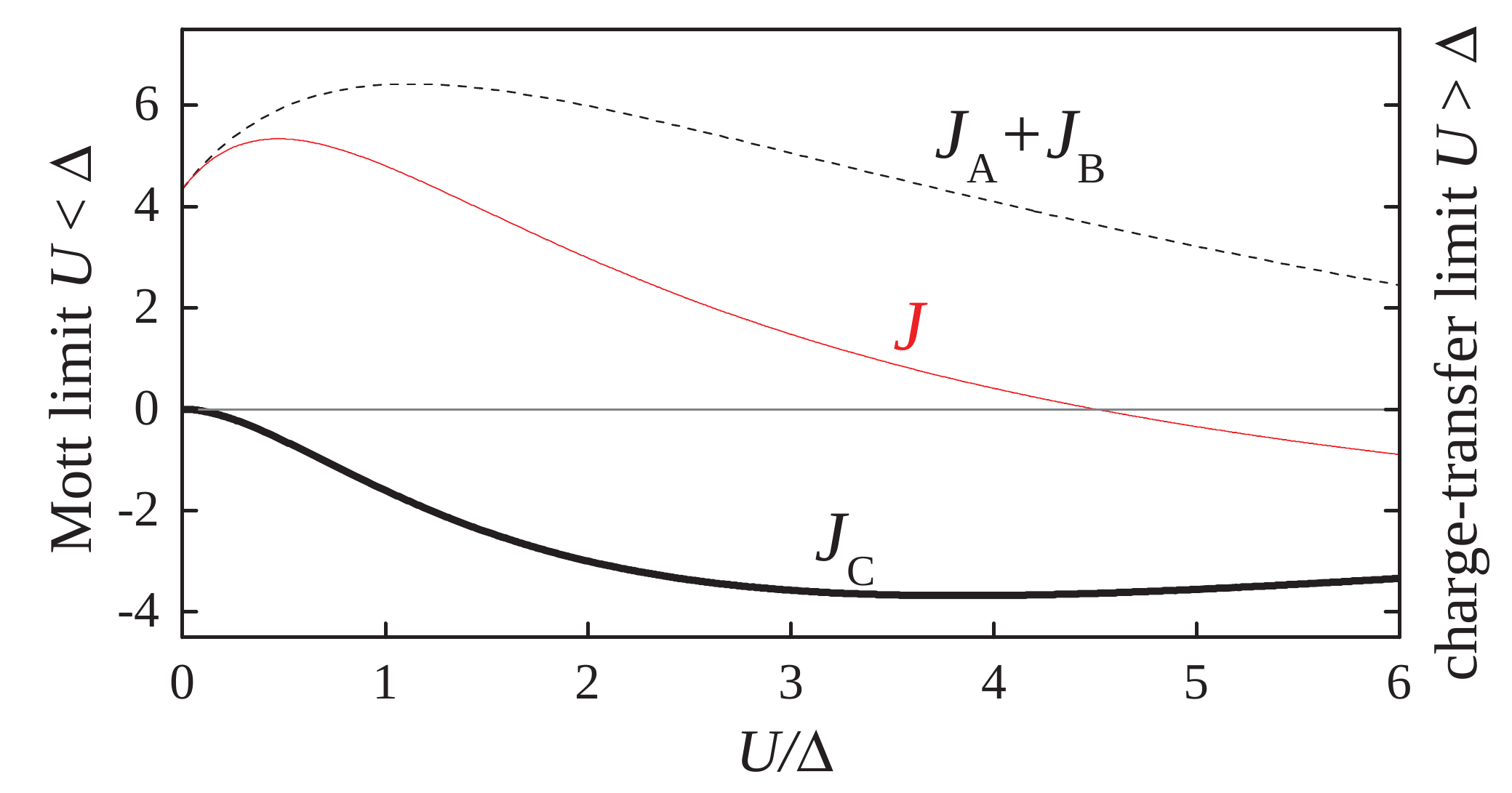}
\caption{Ferromagnetic (of negative sign) Heisenberg coupling $J_{\mathrm{C}}$
  from $e_g-e_g$ hopping process, in units of $t^2/U$, as a function of
  $U/\Delta$. In a charge-transfer regime of large $U/\Delta$, $J_{\mathrm{C}}$
  nearly compensates antiferromagnetic $J_{\mathrm{A}} + J_{\mathrm{B}}$
  contribution (dashed line) from $t_{2g}-t_{2g}$ and $t_{2g}-e_g$ hoppings,
  resulting in strong suppression and sign-change of overall Heisenberg
  coupling $J$ (red thin line). The parameters used are:
  $J^p_H/U_p$=0.3 \cite{Foy13}, $t_{pd\pi}/t_{pd\sigma}=2$, $D/U=0.2$,
  and $U_p/U=0.7$.
}
\label{fig:Jc}
\end{center}
\end{figure}

After calculations following subsection-A2 above, we find the
$e_g-e_g$ charge-transfer contribution to the exchange Hamiltonian:
\begin{equation}
\mathcal{H}^{(c)}_{\mathrm{C}}=-\frac{4}{9}
\frac{t^2_e \; J^p_H}{(\Delta_e+\frac{U'_p}{2})^2}\;\vc S_i \cdot \vc S_j.
\label{eq:c1}
\end{equation}
This is a pure-spin interaction, since both $e_g$ orbitals are half-filled
(no $e_g$ orbital degeneracy). Consequently, after projecting onto pseudospin
subspace, the interaction preserves its SU(2) invariant Heisenberg form:
\begin{align}
\mathcal{H}^{(c)}_{\mathrm{C}} =
J_{\mathrm{C}}\;\vc{\widetilde{S}}_i \cdot \vc{\widetilde{S}}_j,
\end{align}
with
\begin{align}
J_{\mathrm{C}} = -\frac{100}{81}\frac{t^2_e \; J^p_H}{(\Delta_e
+\frac{U'_p}{2})^2}\;.
\label{eq:JC}
\end{align}

In Fig.~\ref{fig:Jc}, we plot $J_{\mathrm{C}}$ as a function of $U/\Delta$,
using the same representative parameters as in Figs.~\ref{fig:kaja}
and~\ref{fig:kbjb} above. For comparison, we show also Heisenberg AF coupling
$J_A+J_B$ originating from $t_{2g}-t_{2g}$ and $t_{2g}-e_g$ channels, as well
as total value of $J$. It follows that the FM exchange coupling
$J_{\mathrm{C}}$ largely compensates the AF contributions of other channels,
as one goes from Mott limit to charge-transfer regime of large $U/\Delta$.

\section{Overall values of J and K: Interplay between different exchange
  mechanisms}
\label{sec:dis}

Having quantified the basic exchange channels for $d^7$ ions, we are now
in position to put the results together:
\begin{equation}
\mathcal{H}^{(c)}_{ij}=J\vc{\widetilde{S}}_i \cdot \vc{\widetilde{S}}_j
+K\widetilde{S}^z_i\widetilde{S}^z_j
+\Gamma(\widetilde{S}^x_i\widetilde{S}^y_j+\widetilde{S}^y_i\widetilde{S}^x_j),
\label{eq:toH}
\end{equation}
with coupling constants
\begin{align}
J &= J_{\mathrm{A}}+J_{\mathrm{B}}+J_{\mathrm{C}}, \notag \\
K &= K_{\mathrm{A}}+K_{\mathrm{B}}, \notag \\
\Gamma &= \Gamma_{\mathrm{A}}.
\label{eq:JKG}
\end{align}
The explicit expressions for individual A($t_{2g}-t_{2g}$), B($t_{2g}-e_g$),
and C($e_g-e_g$) contributions to the exchange constants are given by
Eqs.~(\ref{eq:JKA}), (\ref{eq:JKB}), and (\ref{eq:JC}), respectively.
Using these equations, which constitute the main results of the present work,
one can readily evaluate the overall values of the Hamiltonian parameters $J$,
$K$, and $\Gamma$, and obtain their dependence on material specific parameters
such as $\Delta$, $D$, $J^p_H$, etc.

\begin{figure}
\begin{center}
\includegraphics[width=8.5cm]{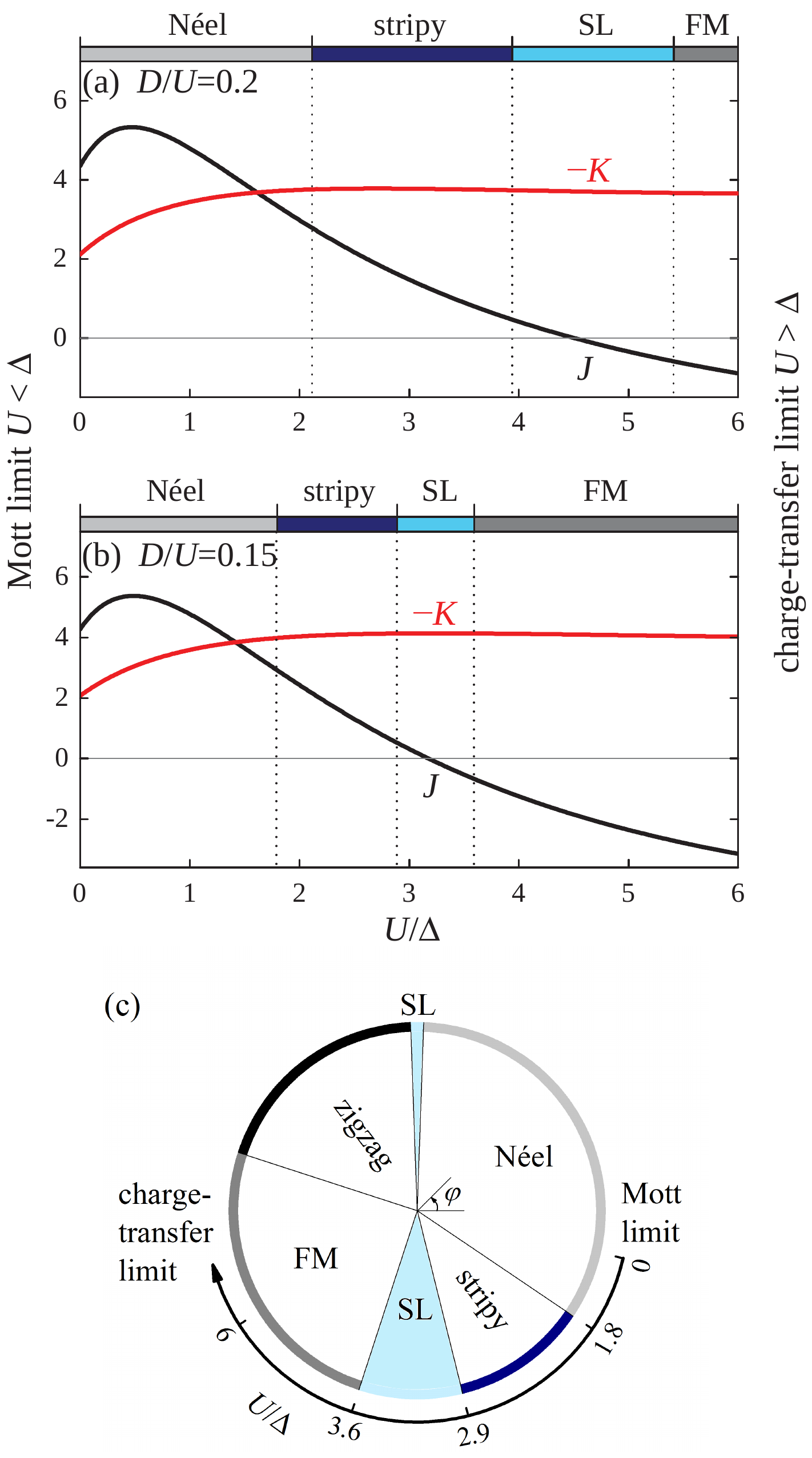}
\caption{Total values of Heisenberg ($J$) and Kitaev ($-K$) couplings
in units of $t^2/U$ as a function of $U/\Delta$, calculated with (a)
$D/U$=0.2 and (b) $D/U$=0.15. Other parameters used are:
$J^p_H/U_p$=0.3 \cite{Foy13}, $t_{pd\pi}/t_{pd\sigma}=2$, $U_p/U=0.7$, and
$\kappa=0.2$. Phase boundaries are obtained using the results of
Refs.~\onlinecite{Cha10,Cha13}.
(c) Projection of the phase boundaries shown in panel (b) onto the phase
diagram of the Kitaev-Heisenberg model in the $\varphi$-angle
representation of Eq.~(\ref{eq:phi}).}
\label{fig:kj}
\end{center}
\end{figure}

As an example, we show in Figs.~\ref{fig:kj}(a) and \ref{fig:kj}(b) the Heisenberg and
Kitaev-type couplings as a function of $U/\Delta$, calculated for two
different values of parameter $D$. We recall that $D$ is ''an effective
$10Dq$'' value in the context of charge-transfer physics, that is a
difference between $p \rightarrow e_g$ and $p \rightarrow t_{2g}$
charge-transfer gaps, renormalized by excitonic
effects, see Eq.~(\ref{eq:D}). In both panels, the Kitaev coupling is always
negative and its value tends to gradually increase with $U/\Delta$. Most
important observation is that Heisenberg coupling $J$ is strongly suppressed and
changes its sign in the charge-transfer regime of large $U/\Delta$, due to
enhanced FM coupling between $e_g$ spins as found above. The resulting
spin-liquid window with $|J| \ll -K$ shifts towards lower values of $U/\Delta$,
when a difference $D$ between $t_{2g}$ and $e_g$ charge-transfer gaps
decreases. Physically, this parameter depends on material properties such as
degree of covalency, crystal structure, etc. For example, from the optical
absorption data in CoO~\cite{Pra59}, we infer $D\sim 1$~eV;
with the ab initio estimates $U\sim 5.0-7.8$~eV~\cite{Ani91,Pic98,Jia10}
this gives $D/U\sim 0.13 - 0.20$.

Regarding $U/\Delta$ parameter in cobalt compounds, this may vary broadly
depending on material chemistry, in particular on the electronegativity of
the anions. While $\Delta \sim 4$~eV in oxides, this value is much reduced in
compounds with Cl, S, P, etc.~\cite{Zaa85,Bre86}, so that
$\Delta \sim 2 - 4$~eV and $U/\Delta \sim 2 - 3$ values seem to be plausible
in cobalt compounds. Given that the Hund's coupling effects shift
the Kitaev spin-liquid window towards much lower values of
$U/\Delta \sim 1 - 3$ in the phase diagram (see Appendix), the
charge-transfer type cobalt insulators may indeed realize the
Kitaev-type interactions that dominate over isotropic Heisenberg couplings.

The $\Gamma$-coupling in Eq.~(\ref{eq:toH}) is contributed by $t_{2g}-t_{2g}$
process~(\ref{eq:JKA}) only, and small when a direct hopping $t'$ is weak
as compared to oxygen-mediated $t$ hopping. The phase diagram of $J-K$ model
alone is often presented in literature using the following
$\varphi$-angle parametrization~\cite{Cha13}:
\begin{align}
\mathcal{H}^{(c)}_{ij}=
A(2\;\mathrm{sin}\varphi\;\widetilde{S}^z_i\widetilde{S}^z_j
+\mathrm{cos}\varphi\;\vc{\widetilde{S}}_i \cdot \vc{\widetilde{S}}_j).
\label{eq:phi}
\end{align}
The energy scale $A$ and angle $\varphi$ are given by
$J=A~\mathrm{cos}\varphi$ and $K=2A~\mathrm{sin}\varphi$. For completeness, we
use this parametrization and map the results of Fig.~\ref{fig:kj}(b) onto
the $\varphi$-angle phase diagram, see Fig.~\ref{fig:kj}(c). This figure
quantifies the phase behavior of $d^7$ pseudospin-1/2 system as a function of
$U/\Delta$, evolving from N\'{e}el AF in Mott limit to FM state in
charge-transfer regime, through an intermediate Kitaev spin-liquid phase.

Recently, new cobalt compounds Na$_2$Co$_2$TeO$_6$ and Na$_3$Co$_2$SbO$_6$ with
a nearly perfect honeycomb lattice of magnetic Co$^{2+}$ $d^7$ ions have been
synthesized and studied~\cite{Vic07,Lef16,Ber17,Won16}.
Both these two systems develop a zigzag-type antiferromagnetic order at low
temperatures, analogous to that observed in $d^5$ pseudospin-1/2
materials RuCl$_3$ and Na$_2$IrO$_3$. The zigzag-AF order can be stabilized
within the Kitaev-Heisenberg model with $K>0$~\cite{Cha13}, or with the help
of longer-range Heisenberg couplings~\cite{Kim11,Cho12} if $K<0$. Since we
found above that the sign of Kitaev coupling in $d^7$ pseudospin-1/2
cobaltates is robustly negative for any values of $U/\Delta$, it seems that
the zigzag-type AF in Na$_2$Co$_2$TeO$_6$ and Na$_3$Co$_2$SbO$_6$ is supported
by $K<0$ and long-range $J$ couplings. As a side remark, we notice that the
long-range pseudospin interactions are expected to be predominantly isotropic,
since many exchange paths are involved and thus bond-directional nature of
orbitals (leading to Kitaev-type interactions at short, nearest-neighbor
distances) can be effectively averaged out at long distances.

In this work, we assumed an ideal cubic symmetry of the pseudospin-1/2
wave functions. Trigonal distortions, likely to be present in real materials,
can induce additional anisotropic terms in the Hamiltonian and affect the phase
boundaries. Neutron and resonant x-ray scattering measurements similar to
those done in RuCl$_3$ and Na$_2$IrO$_3$ (see, e.g.,
Refs.~\onlinecite{Cho12,Chu15,Ban16}) are necessary to quantify the exchange
parameters in $d^7$ pseudospin-1/2 cobalt compounds. Useful information on
anisotropic exchange terms can be deduced also from the analysis of magnetic
anisotropy~\cite{Cha15,Cha16,Siz16} and magnetization~\cite{Jan17} data.

\section{Conclusions}
\label{sec:con}

In this paper, we have presented a comprehensive study of the spin-orbital
exchange interactions between $d^7$ ions with $t^5_{2g}e^2_g$ electronic
configuration. Various exchange channels, involving $t_{2g}$ and $e_g$
orbital interactions in 90$^{\circ}$ bonding geometry, have been examined in
detail and quantified. The exchange processes considered here are generic
to many transition metal compounds, in particular when both $t_{2g}$ and $e_g$
orbitals are spin active.

In a cubic crystal field, the $d^7$ ions with $S=3/2$ and effective orbital
momentum $L=1$ form a pseudospin-1/2 ground state. We have projected
spin-orbital interactions onto this doublet, and obtained the Kitaev-Heisenberg
model as a low-energy magnetic Hamiltonian, with the dominant $K$ term in case
of charge-transfer insulating regime. This is in contrast to $d^7$
cobaltates with 180$^{\circ}$ bonding geometry such as KCoF$_3$, where
isotropic Heisenberg coupling $J$ dominates pseudospin-1/2
interactions~\cite{Hol71,Buy71}, just as in $d^5$ pseudospin-1/2
perovskite Sr$_2$IrO$_4$~\cite{Kim12}.

In $d^5$ compounds such as Na$_x$CoO$_2$, RuCl$_3$, and Na$_2$IrO$_3$,
a suppression of Heisenberg $J$ coupling is due to cancellation of
the $t_{2g}-t_{2g}$ channel $U$ processes in 90$^{\circ}$ bonding exchange
geometry~\cite{Kha05,Jac09}. In $d^7$ cobaltates, we found instead that a
suppression of $J$ coupling is
due to ferromagnetic spin exchange of $e_g$ electrons, which may largely
compensate the AF contribution of other channels or even change the sign of $J$.
This mechanism of $J$-suppression in favor of Kitaev term requires a closeness
to charge-transfer insulating regime. Since cobalt oxides typically belong
to this category of insulators~\cite{Zaa85}, zigzag-AF order observed in
honeycomb lattice cobaltates Na$_2$Co$_2$TeO$_6$ and Na$_3$Co$_2$SbO$_6$ is
likely to have the same origin as in RuCl$_3$ and Na$_2$IrO$_3$, that is, due
to FM Kitaev-type interactions combined with long-range $J$ couplings.

Apart from honeycomb lattice
compounds~\cite{Vic07,Lef16,Ber17,Won16,Bre86,Dav13,Ros17a},
there are many $d^7$ cobaltates possessing pseudospin-1/2 ground state,
such as quasi-one dimensional CoNb$_2$O$_6$~\cite{Col10}, triangular lattice
antiferromagnets Ba$_3$CoSb$_2$O$_9$~\cite{Zho12} and
Ba$_8$CoNb$_6$O$_{24}$~\cite{Raw17}, spinel GeCo$_2$O$_4$~\cite{Tom11} and
pyrochlore lattice NaCaCo$_2$F$_7$~\cite{Ros16,Ros17} cobaltates.
Even though the bonding geometries are no longer exactly 90$^{\circ}$,
a strongly anisotropic, bond-dependent Ising interactions as in $d^5$
(or in $f$-electron~\cite{Li17}) systems are expected in these materials.

Altogether, the results presented in this work suggest that cobalt based
compounds are of interest in the context of pseudospin-1/2 magnetism in
general, and Kitaev model physics in particular, and, as such, they deserve
more focused experimental studies.

\emph{Note added}. Recently, we became aware that a
similar idea has been proposed independently by Sano {\it et al.}~\cite{San17}.


\acknowledgments

We would like to thank J.~Chaloupka for useful discussions. We acknowledge
support by the European Research Council under Advanced
Grant No. 669550 (Com4Com).


\appendix

\section {Hund's coupling effects}

The energies and wavefunctions of the intermediate states, created
during the exchange processes, are affected by intraionic Hund's coupling.
This leads to corrections of the order of $J_H/U$ to the exchange
Hamiltonians (see, e.g., Ref.~\cite{Kha05}). Below, we consider the Hund's
coupling effects on $d^7$ pseudospin-1/2 interactions, and show that they
tend to suppress AF coupling $J$ and hence further support the Kitaev
spin-liquid regime.

A minimal model for the on-site Coulomb and exchange interactions can be
cast in the following form:
\begin{align}
\mathcal{H}_{\mathrm{loc}} &\!=\! U\sum_{i,\alpha}n_{i\alpha\uparrow}n_{i\alpha\downarrow}
\!+\!\sum_{i,\alpha<\beta}\!\left(U'\!-\!J_H \hat{P}_s
\right)n_{i\alpha}n_{i\beta} \notag \\
&+J_H\sum_{i,\alpha\neq\beta} d_{\alpha\uparrow}^\dagger d_{\alpha\downarrow}^\dagger
d_{\beta\downarrow} d_{\beta\uparrow} \;.
\label{eq:JHH}
\end{align}
Here, $\hat{P}_s=(2\vc s_{i\alpha} \vc s_{i\beta}\!+\!\tfrac{1}{2})$ is spin
permutation operator, $n_\alpha$ and $\vc s_\alpha$ are the density and spin-1/2
operators on $\alpha$-orbital, correspondingly. The last, so-called
``pair-hopping'' term describes a motion of doubly occupied orbital states.
$U$ and $U'=U-2J_H$ stand for intraorbital and interorbital Coulomb repulsions,
correspondingly (the spherical symmetry assumed), and $J_H=3B+C \simeq 8B$
is Hund's coupling expressed in terms of Racah parameters $B$ and $C$.
The ratio $J_H/U$ is subject to various screening effects and thus material
sensitive. The representative values of $J_H\simeq 0.8$~eV (from optical data
in CoO~\cite{Pra59}) and $U\sim 5.0-7.8$~eV~\cite{Ani91,Pic98,Jia10} seem to
suggest the range of $0.1< J_H/U < 0.2$, roughly.

Two different valence states of Co ion, $d^6$ and $d^8$, can appear in the
intermediate states. Fortunately, $d^8$ states $t_{2g}^6e_g^2$ and
$t_{2g}^5e_g^3$ are created by hopping processes always in the unique $S=1$
state. Thus, we need to consider Hund's splitting of the $d^6$
intermediate states which appear in the Mott-Hubbard-type $U$ processes
$d^7_id^7_j \rightarrow d^6_id^8_j$ only. The transition energies
$E=E(d^6_id^8_j)-E(d^7_id^7_j)$ and matrix elements depend on the spin-orbital
structure of the initial and intermediate states and hence on Hund's coupling.

\subsection {$t_{2g}$-$t_{2g}$ exchange}

The $t$-hopping generates $d^6$ configuration either in the high-spin $S=2$
state with corresponding excitation energy $E_1=U-3J_H$, or in the low-spin
$S=1$ states at energies $E_2=U+J_H$ and $E_3=U+4J_H$. (We note that
pair-hopping term is essential for obtaining the $S=1$ state energies
and wave functions).

For simplicity, we neglect $J_H/U$ corrections to small $t'$ hopping processes
(as in the main text, we will show the results for $\kappa=t'/t=0.2$), and
focus on oxygen-mediated $t$ hoppings. After somewhat tedious but
straightforward calculations, the exchange contribution
$\mathcal{H}^{(c)}_{\mathrm{A}1}$ from $t_{2g}$-$t_{2g}$ hoppings $\propto t^2/U$
is obtained as follows:
\begin{align}
&\frac{4t^2}{9}\frac{1}{E_1}(\vc S_i \cdot \vc S_j+S^2)
(a_i^{\dag}b_ia_j^{\dag}b_j+b_i^{\dag}a_ib_j^{\dag}a_j)  \notag \\
+&\frac{4 t^2}{27} \left(\frac{1}{E_3}+\frac{2}{E_2}\right)
(\vc S_i \cdot \vc S_j+S^2)(n_{ia}n_{jb}+n_{ib}n_{ja})  \notag \\
-&\frac{t^2}{6}\left(\frac{1}{E_1}\!-\!\frac{1}{E_2}\right)
(\vc S_i \cdot \vc S_j\!+\!S^2)
\left[(n_{ia}\!-\!n_{jb})^2\!+\!(n_{ib}\!-\!n_{ja})^2\right]\notag \\
-&\frac{4 t^2}{27}\left(\frac{1}{E_2}-\frac{1}{E_3}\right)
(\vc S_i \cdot \vc S_j-S^2)
(a_i^{\dag}b_ib_j^{\dag}a_j+b_i^{\dag}a_ia_j^{\dag}b_j)  \notag \\
+&\frac{t^2}{6}\left(\frac{3}{E_1}+\frac{1}{E_2}-\frac{4}{E_3}\right)
(n_{ia}n_{jb}+n_{ib}n_{ja}).
\label{eq:JHA1}
\end{align}
When $J_H=0$, i.e. $E_n=U$, Eq.~\ref{eq:JHA1} fully recovers the first line
($\propto \frac{t^2}{U}$ term) of Eq.~\ref{eq:A11}.

\begin{figure}
\begin{center}
\includegraphics[width=8.5cm]{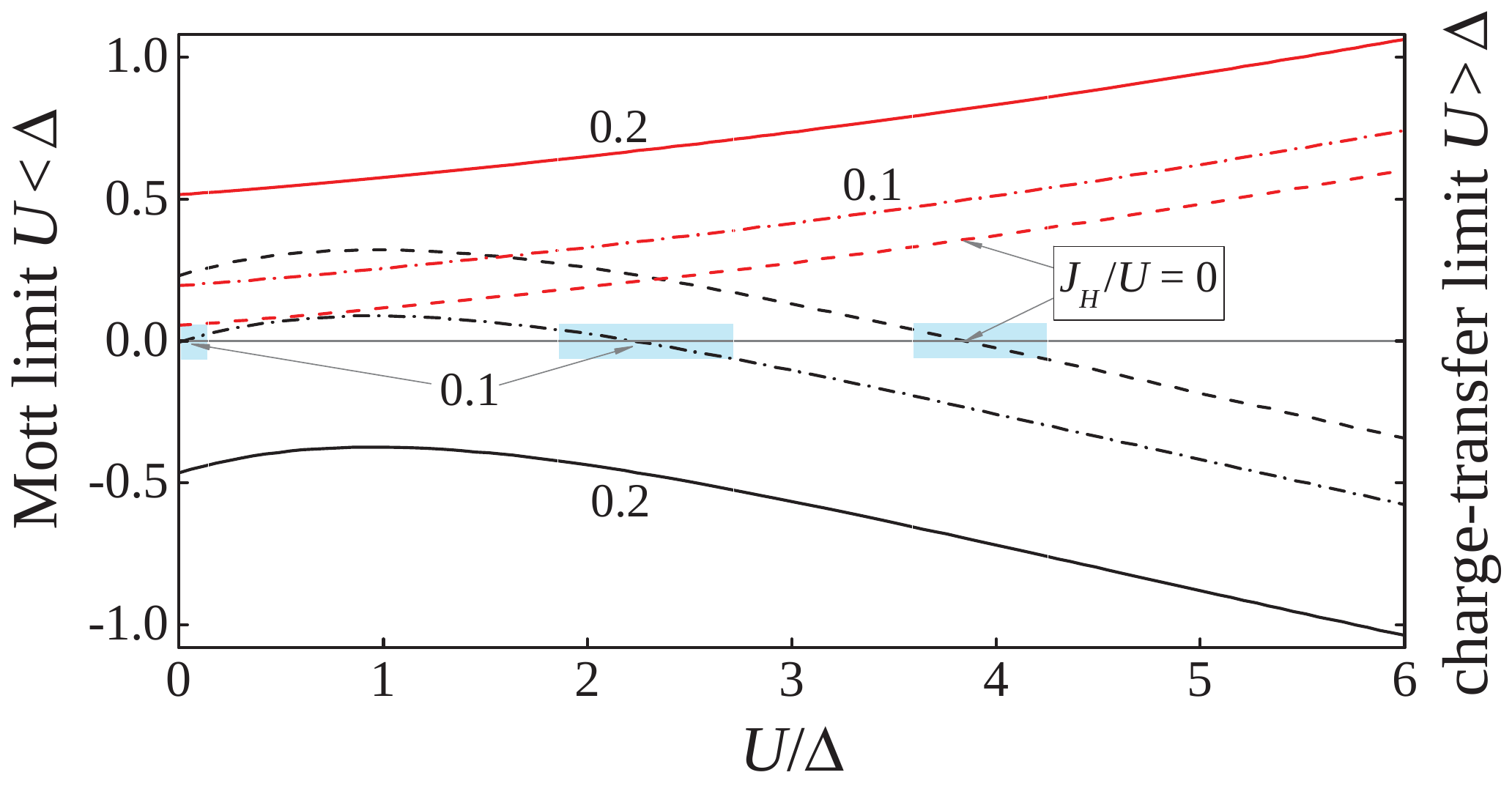}
\caption{$t_{2g}$-$t_{2g}$ contribution to Heisenberg $J_{\mathrm{A}}$ (black
lines) and Kitaev $-K_{\mathrm{A}}$ (red lines) couplings in units of $t^2/U$ as
a function of $U/\Delta$, calculated for different values of Hund's coupling:
$J_H/U=0$ (dashed), $J_H/U=0.1$ (dash-dotted), and $J_H/U=0.2$ (solid lines).
The other parameters (and hence $J_H=0$ results) are the same as in
Fig.~\ref{fig:kaja}. Spin-liquid windows (colored rectangles) are obtained
using the results of Refs.~\onlinecite{Cha10,Cha13}.
}
\label{fig:JHA1}
\end{center}
\end{figure}

After projection onto pseudospin-1/2 doublet, this Hamiltonian
reads as
\begin{align}
\mathcal{H}^{(c)}_{\mathrm{A}1}=J^{(c)}_{\mathrm{A}1}
\vc{\widetilde{S}}_i \cdot \vc{\widetilde{S}}_j
+K^{(c)}_{\mathrm{A}1} \widetilde{S}^z_i\widetilde{S}^z_j,
\end{align}
with the exchange parameters:
\begin{align}
J^{(c)}_{\mathrm{A}1}&=+\frac{t^2}{81}\left(-\frac{31}{E_1}
+\frac{43}{E_2}+\frac{6}{E_3}\right),  \notag \\
K^{(c)}_{\mathrm{A}1}&=-\frac{t^2}{81}\left(\frac{23}{E_1}
-\frac{61}{3E_2}+\frac{4}{3E_3}\right).
\label{eq:JHA2}
\end{align}
In the limit of $J_H=0$ (set $E_n=U$), this gives
$J^{(c)}_{\mathrm{A}1}=\frac{2}{9}\frac{t^2}{U}$ and
$K^{(c)}_{\mathrm{A}1}=-\frac{4}{81}\frac{t^2}{U}$, reproducing the results
$\propto t^2/U$ of the main text (see first line of Eq.~\ref{eq:A12}).

We now replace the first terms ($\propto t^2/U$) of $J_{\mathrm{A}}$
and $K_{\mathrm{A}}$ in Eq.~\ref{eq:JKA} by the corresponding results
from Eq.~\ref{eq:JHA2} (the charge-transfer and cyclic exchange terms
$\propto t^2/\Delta$ are not affected by Hund's coupling). The results
presented in Fig.~\ref{fig:JHA1} show that Hund's coupling corrections
suppress the AF $J$ interaction, which eventually becomes FM for all $U/\Delta$
values when $J_H/U$ is large enough. For the intermediate values of
$J_H/U\sim 0.1$, the spin-liquid window appears around 
$U/\Delta \sim 2$ as well as in the Mott limit.

At this point, we recall that the $t_{2g}$-$t_{2g}$ contribution is 
in fact much weaker than the $t_{2g}$-$e_g$ exchange terms in $d^7$ 
systems. Therefore, Hund's coupling affects $J$ and $K$ parameters 
predominantly via the latter channel,
as shown below.

\begin{figure}
\begin{center}
\includegraphics[width=8.5cm]{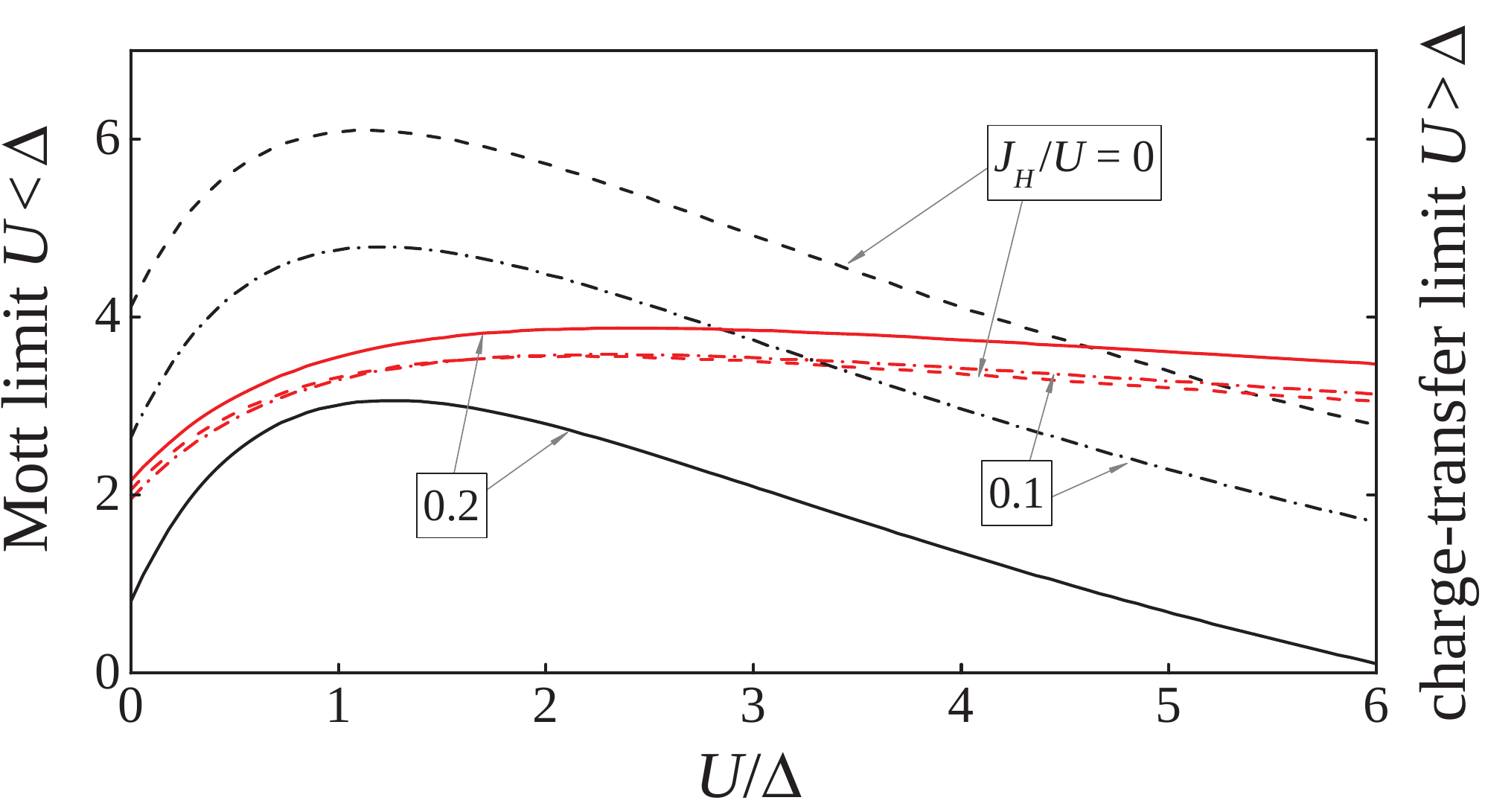}
\caption{$t_{2g}$-$e_g$ contribution to Heisenberg $J_{\mathrm{B}}$ (black lines)
and Kitaev $-K_{\mathrm{B}}$ (red lines) couplings in units of $t^2/U$ as
a function of $U/\Delta$, calculated for different values
of Hund's coupling: $J_H/U=0$ (dashed), $J_H/U=0.1$ (dash-dotted),
and $J_H/U=0.2$ (solid lines). The other parameters (and hence $J_H=0$
results) are the same as in Fig.~\ref{fig:kbjb}.
}
\label{fig:JHA2}
\end{center}
\end{figure}

\begin{figure}
\begin{center}
\includegraphics[width=8.5cm]{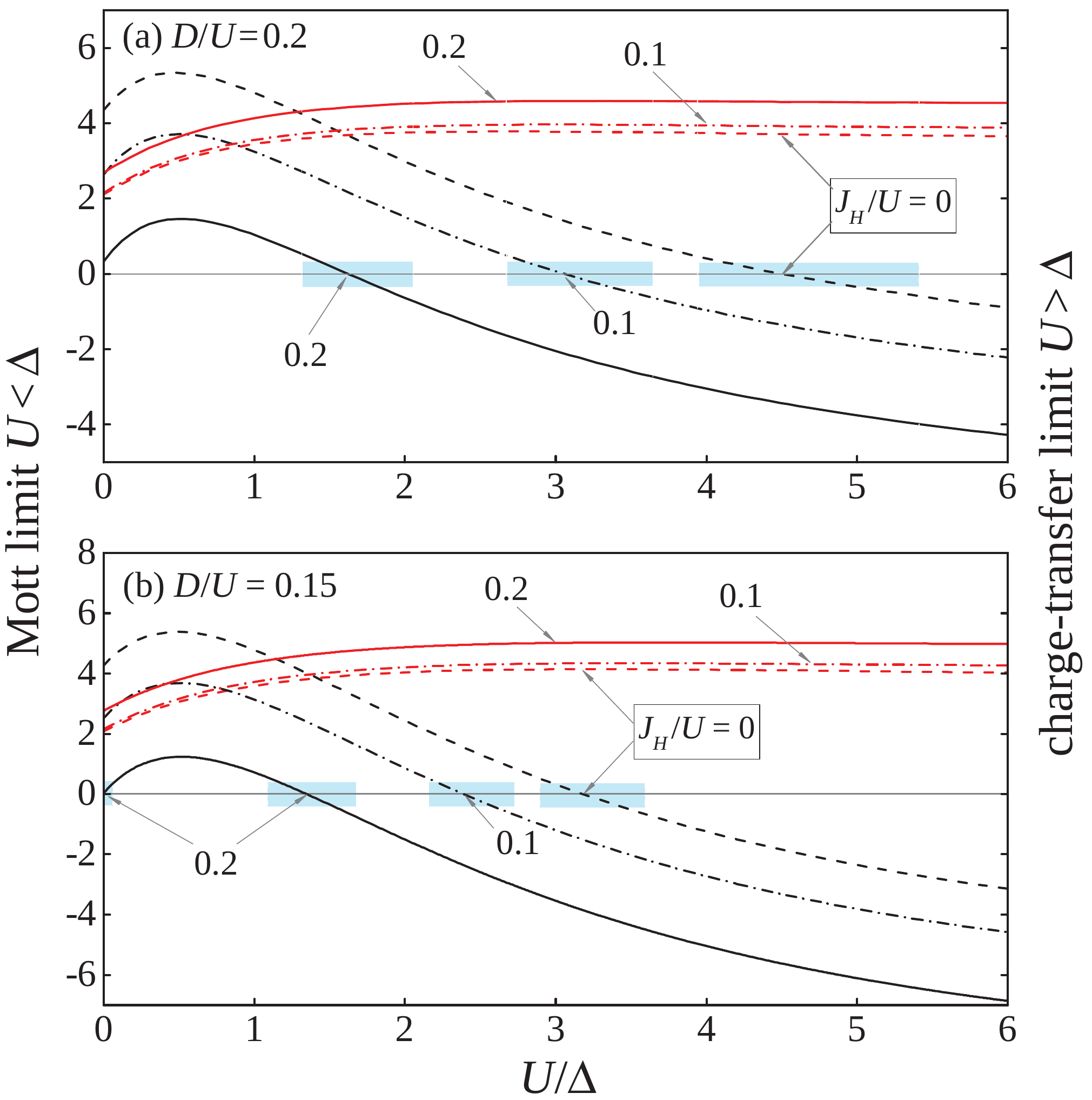}
\caption{Total values of Heisenberg $J$ (black lines) and Kitaev $-K$
(red lines) couplings in units of $t^2/U$ as a function of $U/\Delta$,
calculated with (a) $D/U$=0.2 and (b) $D/U$=0.15 at different Hund's couplings:
$J_H/U=0$ (dashed), $J_H/U=0.1$ (dash-dotted), and $J_H/U=0.2$ (solid lines).
The other parameters (and hence $J_H=0$ results) are the same as in
Fig.~\ref{fig:kj}. Spin-liquid windows (colored rectangles) are obtained
using the results of Refs.~\onlinecite{Cha10,Cha13}. For $J_H/U=0.2$, small
spin-liquid phase appears also in the Mott limit of $U/\Delta=0$.}
\label{fig:JHA3}
\end{center}
\end{figure}

\subsection {$t_{2g}$-$e_g$ exchange}
The $t_{2g}$-$e_g$ hoppings generate $d^6$ intermediate states with both high
$S=2$ and low $S=1$ spins, too, but the energies $E_{1,2,3}$ defined above are
now shifted by splitting $D$ between $t_{2g}$ and $e_g$ orbitals. Collecting
all possible transitions with the corresponding matrix elements, we find that
at finite Hund's coupling the Hamiltonian $\mathcal{H}^{(c)}_{\mathrm{B}1}$
in Eq.~\ref{eq:B1} takes the following form:
\begin{align}
\mathcal{H}^{(c)}_{\mathrm{B}1} &=
\frac{4\alpha}{9}\frac{tt_e}{\widetilde{U}}
(\vc S_i \cdot \vc S_j-S^2)(n_{ic}+n_{jc})\notag \\
&-\frac{tt_e}{6}\frac{\Delta_e}{\Delta}
\left(\frac{1}{E_1+D}\!-\!\frac{1}{E_2+D} \right)
\vc S_i \cdot \vc S_j (n^i_{ab}+n^j_{ab}),
\label{eq:JHB1}
\end{align}
with the renormalized constants $\alpha$ and $1/\widetilde{U}$:
\begin{align}
\alpha &= 1-\frac{D^2}{2\Delta\Delta_e}
\left(\frac{\Delta+\Delta_e}{U+2J_H}-1\right),\notag \\
\frac{1}{\widetilde{U}} &= \frac{1}{6}\left(\frac{2}{E_2+D}\!+\!
\frac{1}{E_3+D}\!+\!\frac{3}{U+2J_H-D}\right).
\end{align}
We recall that $n_{ab}=n_a+n_b=1-n_c$ (in hole representation). Note also that
new transition energy $U+2J_H-D$ appeared in $1/\widetilde{U}$. This
corresponds to $d^6(t_{2g}^5e_g^1)$ state with two holes on an $e_g$ orbital;
its pair-hopping motion to the $t_{2g}$ levels is suppressed by $2D > J_H$
splitting.

After projection onto pseudospin-1/2 doublet, the Hamiltonian reads as
\begin{align}
\mathcal{H}^{(c)}_{\mathrm{B}1}=J^{(c)}_{\mathrm{B}1}
\vc{\widetilde{S}}_i \cdot \vc{\widetilde{S}}_j
+K^{(c)}_{\mathrm{B}1} \widetilde{S}^z_i\widetilde{S}^z_j,
\end{align}
with exchange parameters:
\begin{align}
J^{(c)}_{\mathrm{B}1}&=+\frac{80}{81}tt_e \left[
\frac{\alpha}{\widetilde{U}}-\frac{9\Delta_e}{16\Delta}
\left(\frac{1}{E_1+D}-\frac{1}{E_2+D} \right) \right],  \notag \\
K^{(c)}_{\mathrm{B}1}&=-\frac{40}{81}tt_e \left[
\frac{\alpha}{\widetilde{U}}+\frac{3\Delta_e}{8\Delta}
\left(\frac{1}{E_1+D}-\frac{1}{E_2+D} \right) \right].
\label{eq:JHB2}
\end{align}
These equations tell that Hund's coupling tends to reduce AF $J$ and increase
FM $K$ values [these corrections originate from the last term in
Eq.~\ref{eq:JHB1}]. When $J_H=0$ (set $E_n=U$), these results recover
Eq.~\ref{eq:b1} of the main text.

In Fig.~\ref{fig:JHA2}, we show the $t_{2g}$-$e_g$ hopping contribution to
$J_{\mathrm{B}}$ and $K_{\mathrm{B}}$ values at different $J_H/U$ ratios,
including the $J_H=0$ results of Fig.~\ref{fig:kbjb} for comparison. While
$K$ value remains nearly unaffected, Hund's coupling strongly suppresses
AF $J$ coupling, which, however, remains positive even at $J_H/U=0.2$.

Having considered Hund's coupling effects both in the $t_{2g}$-$t_{2g}$ and
$t_{2g}$-$e_g$ channels ($e_g$-$e_g$ channel remains unchanged), we are ready
to discuss the overall values of $J$ and $K$ parameters. Putting the results
together into Eq.~\ref{eq:JKG}, we obtain the exchange parameters shown in
Fig.~\ref{fig:JHA3} (including the $J_H=0$ results of Fig.~\ref{fig:kj}).
As expected, the major effect of Hund's coupling is to shift down the
$J$ curves, such that Heisenberg coupling changes its sign at the lower values
of $U/\Delta$ ratio. This implies that Hund's coupling cooperates with
the charge-transfer effects to support Kitaev spin-liquid regime of
$J/K\sim 0$. We observe that, at $J_H/U=0.2$, this regime may appear also
in the Mott limit, consistent with the recent work by
Sano {\it et al.}~\cite{San17}.

Overall, it seems that Hund's coupling effects in $d^7$ pseudospin-1/2
compounds shift the spin-liquid parameter regime towards the lower values
of $U/\Delta$ ratio, thereby increasing the chances of finding the Kitaev model
physics in a broader class of $d^7$ cobalt compounds.


\end{document}